\documentclass[11pt]{article}
\usepackage[centertags]{amsmath}
\usepackage{amsfonts}
\usepackage{amssymb}
\usepackage{amsthm}
\usepackage{newlfont}

\numberwithin{equation}{section}

\newtheorem{theorem}{Theorem}
\newtheorem{lemma}{Lemma}

\newtheorem{corollary}{Corollary}

\begin{document}
\title{A rigorous Hermitian proof about the G-dynamics and analogy with Berry-Keating's Hamiltonian}
\author{Jack Whongius\thanks{School of Mathematical Sciences,~ Xiamen
University,~ 361005,  China.~ The research of the author is  partially  supported by NSF grants. \newline Email:  fmsswangius@stu.xmu.edu.cn.}}

\maketitle

\par
\par Quantum covariant Hamiltonian system theory provides a coherent framework for modelling the complex dynamics of quantum systems.  In this paper, we centrally deal with the Hermiticity of quantum operators that directly links to the  physical observable, thusly, we give a rigorous proof to verify one-dimensional G-dynamics  ${{\hat{w}}^{\left( cl \right)}}={{\hat{w}}^{\left( cl \right)\dagger }}\in Her$ that is a Hermitian operator satisfying $\left( {{{\hat{w}}}^{\left( cl \right)}}\phi ,\varphi  \right)=\left( \phi ,{{{\hat{w}}}^{\left( cl \right)}}\varphi  \right)$ for any two states $\phi$ and $\varphi$, and its eigenvalues are real. We also prove that curvature operator is a skew-Hermitian operator as well.  The act of finishing this Hermitian proof valuably enables us to ensure the non-Hermitian Hamiltonian operator ${{\hat{H}}^{\left( ri \right)}} ={{\hat{H}}^{\left( g \right)}} -{{\hat{H}}^{\left( \operatorname{clm} \right)}}\in NHer$ that is divided into the Hermitian operator ${{\hat{H}}^{\left( g \right)}} ={{\hat{H}}^{\left( cl \right)}}-{{E}^{\left( s \right)}}/2\in Her$ and the skew-Hermitian operator ${{\hat{H}}^{\left( \operatorname{clm} \right)}}=\sqrt{-1}\hbar {{\hat{w}}^{\left( cl \right)}}\in SHer$ generally, and ${{\hat{H}}^{\left( ri \right)}}$ always has the complex eigenvalues.
  We use the formula of the G-dynamics to evaluate the Berry-Keating's Hamiltonian operator ${{\hat{H}}^{\left( \text{bk}\right)}}=-\sqrt{-1}\hbar \hat{\theta }/2\in Her$  and its extensive version $\hat{H}^{\left( \text{gbk}\right)}\in NHer$ as the applications of the G-dynamics, to see how the similarity appears in the light of obvious factor $\hat{\theta }/2=x\frac{d}{dx}+1/2\in SHer$, etc.

\section*{Notation and terminology}

\par
The following abbreviations are used in this manuscript, we denote by $\ast$ the complex conjugate .
In contrast, $\dag$ symbol represents the conjugate transpose operation which is supposed to be taking the transpose and then taking the complex conjugate of each entry. A tilde $\sim$ over an expression denotes the transpose of an operator; for instance, $\widetilde{\hat{A}}$  is the transpose of operator $\hat{A}$. And $\left\langle~~ \right\rangle$ means the expectation value, a hat $\hat{\bullet}$ over an expression denotes the quantum operator, the rests denote the functions without hat, for example, $\hat{f}$ is a quantum operator of the function $f$. Unless stated otherwise.

We denote by $Her$ the set of Hermitian operators, and the set of non-Hermitian operators is $NHer$, the set of skew-Hermitian operators or anti-Hermitian operators  is denoted by $SHer$ in each argument.
\par$\left[ \hat{f},\hat{g} \right]_{QPB}=\hat{f}\hat{g}-\hat{g}\hat{f}$  denotes the quantum Poisson brackets (QPB) with respect to two operators $\hat{f}, \hat{g}$.
 $\left[ \hat{f},\hat{g} \right]={{\left[ \hat{f},\hat{g} \right]}_{QPB}}+G\left( s,\hat{f},\hat{g} \right)$ is the quantum covariant Poisson bracket (QCPB), and $G\left(s,\hat{f},\hat{g} \right)=\hat{f}{{\left[ s,\hat{g} \right]}_{QPB}}-\hat{g}{{\left[ s,\hat{f} \right]}_{QPB}}$ is the quantum geometric bracket, and geometric scalar function $s$ represents the inherent properties of manifolds, we denote by $\hat{I}\left( \hat{f},\hat{g},s \right)={{\left[ \hat{f},\hat{g} \right]}_{QPB}}-\hat{g}{{\left[ s,\hat{f} \right]}_{QPB}}$ coupling terms with the space manifold, especially, $\hat{I}\left( s,\hat{g},s \right)={{\left[ s,\hat{g} \right]}_{QPB}}$ and $\hat{I}\left( \hat{f},\hat{g},0 \right)={{\left[ \hat{f},\hat{g} \right]}_{QPB}}$.

\section{Introduction and main results}
In the study of dynamic behavior of G-dynamics $\hat{w}=-\sqrt{-1}{{\left[ s,\hat{H} \right]}_{QPB}}/\hbar$ as a part of the covariant dynamics $\frac{\mathcal{D}\hat{f}}{dt}=-\sqrt{-1}\left[ \hat{f},\hat{H} \right]/\hbar$ defined by quantum covariant Hamiltonian system (QCHS) $\left[ \hat{f},\hat{H} \right]=\sqrt{-1}\hbar\frac{\mathcal{D}\hat{f}}{dt}$, one very often comes across the uncertainty of Hermiticity, both analytically and numerically. The difficulties originate from the complexity of the system stemming from the geometry.
Initiating the basic concepts of the G-dynamics is begun since 2020s, as we summarized,
a brief description of the framework of the quantum covariant Poisson bracket (QCPB)  [G-W] theory is $\left[ \hat{f},\hat{g} \right]=\hat{I}\left( \hat{f},\hat{g},s \right)+\hat{f}\hat{I}\left( s,\hat{g},s \right)$ systematically.
\begin{align}
  & \text{QPB}\to \text{QCPB}\to\text{QCHS}\to \text{covariant dynamics} \notag\\
 & \begin{matrix}
   {} & {} & {} & {}  \\
\end{matrix}\begin{matrix}
   {} & {} & {} & {}  \\
\end{matrix}\begin{matrix}
   {} & {} & {} & \begin{matrix}
      {} &{} & {} & {} & {} & {} & {} & \swarrow  & {} & \searrow   \\
\end{matrix}  \\
\end{matrix}  \notag\\
 & \begin{matrix}
   {} & {} & {} & {}  \\
\end{matrix}\begin{matrix}
   {} & {} & {} & \begin{matrix}
   {} & {} & {} & \begin{matrix}
     {} &{} &\text{G-dynamics} & {} & \begin{matrix}
   {} & \text{generalized Heisenberg equation} & {}  \\
\end{matrix}   \notag\\
\end{matrix}   \notag\\
\end{matrix}   \notag\\
\end{matrix} \notag
\end{align}Note that the G-dynamics $\hat{w}$ constructs the quantum covariant time operator $\frac{\mathcal{D}}{dt}=d/dt+\hat{w}$ in which the G-dynamics is $\hat{w}=-\sqrt{-1}\hat{I}\left( s,\hat{H},s \right)/\hbar$  while an observable $\hat{f}$ is a physical quantity that can be measured which satisfies the generalized Heisenberg equation of motion $d\hat{f}/dt =-\sqrt{-1}\hat{I}\left( \hat{f},\hat{H},s \right)/\hbar $ expanded detailedly by 
$d\hat{f}/dt =-\sqrt{-1}\hat{I}\left( \hat{f},\hat{H},0 \right)/\hbar +\sqrt{-1}\hat{H}\hat{I}\left( s,\hat{f},s \right)/\hbar$  
 as a naturally geometric generalization of the Heisenberg equation that $d\hat{f}/dt=-\sqrt{-1}\hat{I}\left( \hat{f},\hat{H},0 \right)/\hbar $ holds in terms of the operator $\hat{f}$ with no explicit time dependence, in such case, the wave function $\psi$ does not change with time.  
Recent developments on one-dimensional case of the G-dynamics have been made by [I] which focuses on the algebraical analysis of two various kinds of one-dimensional G-dynamics  ${{\hat{w}}^{\left( cl \right)}}$ and ${{\hat{w}}^{\left( ri\right)}}$ deduced by various Hamiltonian operator $\hat{H}$ in using the formula $\hat{w}=-\sqrt{-1}{{\left[ s,\hat{H} \right]}_{QPB}}/\hbar$ in which $s$ is a scalar real-valued function on $M$. The G-dynamics  $\hat{w}$ is always expected to be non-zero for a complete quantum theory.  The geometric wave equation $\sqrt{-1}\hbar \hat{w}\psi ={{\hat{H}}^{\left( \operatorname{Im} \right)}}\psi$ follows for a given wave function $\psi$, where we denote ${{\hat{H}}^{\left( \operatorname{Im} \right)}}={{\hat{H}}^{\left( \operatorname{Im} \right)}}\left( \hat{w} \right)$ for simplicity, and ${{\hat{H}}^{\left( \operatorname{Im} \right)}}={{\left[ s,\hat{H} \right]}_{QPB}}=\hat{I}\left( s,\hat{H},s \right)$ is the geometric Hamiltonian operator. [I] has proven a series of new results on two different kinds of G-dynamics including an important identity ${{\hat{w}}^{\left( cl \right)}} {{u}^{-1/2}}=0$, it states that there always exists function ${{u}^{-1/2}}$ such that identity ${{\hat{w}}^{\left( cl \right)}} {{u}^{-1/2}}=0$ holds for all $u>0$.
 Note that the G-dynamics model $${{\hat{w}}^{\left( cl \right)}}=-\sqrt{-1}{{\left[ s,{{\hat{H}}^{\left( cl\right)}}\right]}_{QPB}}/\hbar=-\sqrt{-1}\hat{I}\left( s,{{\hat{H}}^{\left( cl\right)}},s \right)/\hbar$$ admits the following one-dimensional representation
  \begin{align}\label{a1}
    {{\hat{w}}^{\left( cl \right)}}&={{b}_{c}}\left( 2u\frac{d}{dx}+u_{x} \right)=u{{\hat{v}}^{\left( cl \right)}}+\frac{1}{2}{{\hat{v}}^{\left( cl \right)}}u={{b}_{c}}\hat{Q}
  \end{align}
that is induced by the classical Hamiltonian operator ${{\hat{H}}^{\left( cl \right)}}=-\frac{{{\hbar }^{2}}}{2m}\frac{{{d}^{2}}}{d{{x}^{2}}}+V\left( x \right)\in Her$ as a Hermitian operator, where line curvature $u=ds/dx=s_{x}$ and $u_{x}=du/dx$ have been used, and $b_{c}=-\sqrt{-1}\hbar/2m$, the core part $\hat{Q}$ is the curvature operator shown by
$\hat{Q}={{\hat{w}}^{\left( cl \right)}}/{{b}_{c}}=2u\frac{d}{dx}+u_{x}
$ which is satisfying $\hat{Q}{u}^{-1/2}=0$, 
where ${{\hat{v}}^{\left( cl \right)}}={{\hat{p}}^{\left( cl \right)}}/m=2b_{c}d/dx\in Her$ is one dimensional velocity operator,
for this reason, one-dimensional G-dynamics \eqref{a1} as proven is a linear operator and rewritten as
\begin{equation}\label{a4}
  {{\hat{w}}^{\left( cl \right)}}=\left( u{{\hat{p}}^{\left( cl \right)}}+\frac{1}{2}{{\hat{p}}^{\left( cl \right)}}u \right)/m
\end{equation}
or in another form
$m{{\hat{w}}^{\left( cl \right)}}=u{{\hat{p}}^{\left( cl \right)}}+\frac{1}{2}{{\hat{p}}^{\left( cl \right)}}u$, where ${{\hat{p}}^{\left( cl \right)}}$ is the classical momentum operator.  
The geometric wave equation based on \eqref{a1}  follows
\[\sqrt{-1}\hbar {{\hat{w}}^{\left( cl \right)}}\psi ={{\hat{H}}^{\left( \operatorname{clm} \right)}}\psi =\frac{{{\hbar }^{2}}}{2m}\left( 2u{{\psi }_{x}}+\psi {{u}_{x}} \right)\]where we denote
${{\hat{H}}^{\left( \operatorname{clm} \right)}}={{\hat{H}}^{\left( \operatorname{Im} \right)}}\left( {{{\hat{w}}}^{\left( cl \right)}} \right)=\hat{I}\left( s,{{\hat{H}}^{\left( cl\right)}},s \right)$ for convenience.  Another one-dimensional G-dynamics with respect to ${{\hat{H}}^{\left( ri \right)}} $ is given by
\begin{equation}\label{b1}
  {{\hat{w}}^{\left( ri \right)}} =-\sqrt{-1}{{\left[ s,{{\hat{H}}^{\left( ri \right)}} \right]}_{QPB}}/\hbar={{\hat{w}}^{\left( cl \right)}}+{{w}^{\left( s \right)}}
\end{equation}
where the non-Hermitian Hamiltonian operator
\begin{equation}\label{a2}
  {{\hat{H}}^{\left( ri \right)}} ={{{\hat{H}}}^{\left( cl \right)}}-{{E}^{\left( s \right)}}/2-\sqrt{-1}\hbar {{\hat{w}}^{\left( cl \right)}}={{\hat{T}}^{\left( ri \right)}}+V(x)
\end{equation}
is used, and
 ${{w}^{\left( s \right)}}={{\left[ {{\hat{w}}^{\left( cl \right)}},s \right]}_{QPB}}=2{{b}_{c}}{{u}^{2}}$, and ${{E}^{\left( s \right)}}=\frac{{{\hbar }^{2}}}{m}{{u^{2}}}=\sqrt{-1}\hbar{{w}^{\left( s \right)}}$ can be regarded as a potential energy, hence, $${{\hat{H}}^{\left( ri \right)}}={{\hat{H}}^{\left( cl \right)}}-\sqrt{-1}\hbar {{w}^{\left( s \right)}}/2-\sqrt{-1}\hbar {{\hat{w}}^{\left( cl \right)}}$$ we denote  ${{\hat{H}}^{\left( g \right)}} ={{{\hat{H}}}^{\left( cl \right)}}-{{E}^{\left( s \right)}}/2$, then \eqref{a2} is simply given by ${{\hat{H}}^{\left( ri \right)}} ={{\hat{H}}^{\left( g \right)}} -{{\hat{H}}^{\left( \operatorname{clm} \right)}}$ as a non-Hermitian Hamiltonian that arises naturally in quantum systems as effective interactions for a subsystem, where the imaginary geomenergy is
${{\hat{H}}^{\left( \operatorname{clm} \right)}}=\sqrt{-1}\hbar {{\hat{w}}^{\left( cl \right)}}=\hat{I}\left( s,{{\hat{H}}^{\left( cl\right)}},s \right)$.
It turns out to be useful to study the algebraical structure of the one-dimensional G-dynamics, it will open up a spetaculous new insight to understand the bizard quantum world.
The geometrinetic energy operator
$${{\hat{T}}^{\left( ri \right)}}=-\frac{{{\hbar }^{2}}}{2m}\left( {{d}^{2}}/{dx}^{2}+{{u}^{2}} \right)-\sqrt{-1}\hbar {{\hat{w}}^{\left( cl \right)}}$$ has led to  ${{\hat{T}}^{\left( ri \right)}}/\hbar =-\sqrt{-1}{{b}_{c}}{{\hat{Q}}^{\left( c \right)}}$
where ${{\hat{Q}}^{\left( c \right)}}={{d}^{2}}/d{{x}^{2}}+{{u}^{2}}
+\hat{Q}$. Similarly, the geometric wave equation in terms of \eqref{b1} ${{\hat{w}}^{\left( ri \right)}}$ appears
\begin{align}
 \sqrt{-1}\hbar {{{\hat{w}}}^{\left( ri \right)}}\psi & ={{{\hat{H}}}^{\left( \operatorname{rim} \right)}}\psi =\frac{{{\hbar }^{2}}}{2m}\left( 2u{{\psi }_{x}}+\psi {{u}_{x}}+2\psi{{u}^{2}} \right) \notag
\end{align}where we denote ${{{\hat{H}}}^{\left( \operatorname{rim} \right)}}={{{\hat{H}}}^{\left( \operatorname{Im} \right)}}\left( {{{\hat{w}}}^{\left( ri \right)}} \right)=\hat{I}\left( s,{{\hat{H}}^{\left( ri\right)}},s \right)$, and then
${{{\hat{H}}}^{\left( \operatorname{rim} \right)}}\psi={{{\hat{H}}}^{\left( \operatorname{clm} \right)}}\psi +{{E}^{\left( s \right)}}\psi$.   
Note that for the non-Hermitian Hamiltonian operator \eqref{a2}, if $V(x)={{E}^{\left( s \right)}}/2$ is taken for ${{\hat{H}}^{\left( ri \right)}}$, then ${{\hat{H}}^{\left( ri \right)}}$ becomes one-dimensional motor operator
\begin{equation}\label{a7}
  {{\hat{E}}^{\left( w \right)}}=-\frac{{{\hbar }^{2}}}{2m}{d}^{2}/d{{x}^{2}}-\sqrt{-1}\hbar {{\hat{w}}^{\left( cl \right)}}=-\frac{{{\hbar }^{2}}}{2m}{d}^{2}/d{{x}^{2}}-{{\hat{H}}^{\left( \operatorname{clm} \right)}}
\end{equation}
And \eqref{b1} is essentially rewritten as ${{\hat{w}}^{\left( ri \right)}} =-\sqrt{-1}{{\left[ s,{{\hat{E}}^{\left( w \right)}} \right]}_{QPB}}/\hbar$.
By using the notion $\hat{I}\left( \hat{f},\hat{g},s \right)={{\left[ \hat{f},\hat{g} \right]}_{QPB}}-\hat{f}{{\left[ s,\hat{g} \right]}_{QPB}}+G\left( s,\hat{f},\hat{g} \right)$, we can get the following expressions:
\begin{align}
  & {{\hat{w}}^{\left( cl \right)}}=-\sqrt{-1}\hat{I}\left( s,{{\hat{H}}^{\left( cl \right)}},s \right)/\hbar  \notag\\
 & u=-\sqrt{-1}\hat{I}\left( s,{{\hat{p}}^{\left( cl \right)}},s \right)/\hbar  \notag\\
 & {{\hat{w}}^{\left( ri \right)}}=-\sqrt{-1}\hat{I}\left( s,{{\hat{H}}^{\left( ri \right)}},s \right)/\hbar  \notag\\
 & {{u}^{2}}/m=-\sqrt{-1}\hat{I}\left( s,{{\hat{w}}^{\left( cl \right)}},s \right)/\hbar  \notag\\
 & 0=-\sqrt{-1}\hat{I}\left( s,x,s \right)/\hbar =-\sqrt{-1}\hat{I}\left( s,u,s \right)/\hbar  \notag
\end{align}
We get the relations given by
$\hat{I}\left( \hat{f},\hat{g},s \right)-\hat{I}\left( \hat{f},\hat{g},0 \right)=-\hat{g}{{\left[ s,\hat{f} \right]}_{QPB}}$
and the quantum geometric bracket $G\left( s,\hat{f},\hat{g} \right)=\hat{f}\hat{I}\left( s,\hat{g},s \right)+\hat{I}\left( \hat{f},\hat{g},s \right)-\hat{I}\left( \hat{f},\hat{g},0 \right)$. Actually, for quantum geometric bracket in terms of operators $\hat{f},\hat{H} $, it yields
\begin{align}
 G\left( s,\hat{f},\hat{H} \right) &=\sqrt{-1}\hbar \hat{f}\hat{w}-\hat{H}{{\left[ s,\hat{f} \right]}_{QPB}}  =\sqrt{-1}\hbar \left( \hat{f}\hat{w}+\sqrt{-1}\hat{H}{{\left[ s,\hat{f} \right]}_{QPB}}/\hbar  \right) \notag
\end{align}
The eigenvalues equation of ${{\hat{w}}^{\left( cl \right)}}$ reads
${{\hat{w}}^{\left( cl \right)}}\psi ={{w}^{\left( q \right)}}\psi$ for the eigenvalues ${w}^{\left( q \right)}$ as a geometric frequency  of the G-dynamics
${{\hat{w}}^{\left( cl \right)}}$,  $\psi={{C}_{0}}F \left( u,t,{{w}^{\left( q \right)}} \right)$ is the eigenfunction of the G-dynamics ${{\hat{w}}^{\left( cl \right)}}$ that is associated with eigenvalue ${w}^{\left( q \right)}$, where
$F \left( u,t,{{w}^{\left( q \right)}} \right)={{u}^{-1/2}}{{e}^{\sqrt{-1}{{w}^{\left( q \right)}}t}}={{u}^{-1/2}}{{e}^{\sqrt{-1}\frac{{{E}^{\left( q \right)}}}{\hbar }t}}
$, where ${{E}^{\left( q \right)}}=\hbar{{w}^{\left( q \right)}} $ is the geometric energy eigenvalue.
In fact, $L\left( u,t,\sqrt{-1}w^{(q)}  \right)= F \left( u,t,{{w}^{\left( q \right)}} \right)$ holds for all $u>0$. Conclusively, there are obvious properties in terms of the G-dynamics ${{{\hat{w}}}^{\left( cl \right)}}$ given as follows:
\begin{description}
  \item[i] ${{{\hat{w}}}^{\left( cl \right)}}{{u}^{-1/2}}=0$
  \item[ii] ${{{\hat{w}}}^{\left( cl \right)}}L\left( u,t,\sqrt{-1}{{w}^{\left( q \right)}} \right)={{w}^{\left( q \right)}}L\left( u,t,\sqrt{-1}{{w}^{\left( q \right)}} \right)$
  \item[iii] $L\left( \prod\limits_{j}{{{u}_{j}}},\sum\limits_{j}{{{t}_{j}}},\sqrt{-1}{{w}^{\left( q \right)}} \right)=\prod\limits_{j}{L\left( {{u}_{j}},{{t}_{j}},\sqrt{-1}{{w}^{\left( q \right)}} \right)},~~j=1,2,\cdots,\infty$
    \item[iv]  ${{\hat{H}}^{\left( \operatorname{clm} \right)}}L\left( u,t,\sqrt{-1}{{w}^{\left( q \right)}} \right)=\sqrt{-1}{{E}^{\left( q \right)}}L\left( u,t,\sqrt{-1}{{w}^{\left( q \right)}} \right)$
\end{description}
The covariant evolutions of the quantum mechanical coordinate operator $x$ and ${{\hat{p}}}^{\left( cl \right)}$ read
$\frac{\mathcal{D}}{dt}x\left( t \right)={{\hat{v}}^{\left( cl \right)}}+x\left( t \right){{\hat{w}}^{\left( cl \right)}}$, and classical momentum operator
$$\frac{\mathcal{D}}{dt}{{{\hat{p}}}^{\left( cl \right)}}\left( t \right)=-m{{\omega }^{2}}x-\hat{H}^{\left( cl \right)}u+{{\hat{p}}^{\left( cl \right)}}\left( t \right){{\hat{w}}^{\left( cl \right)}}$$ respectively, as it shows in terms of the
covariant evolution of the classical momentum operator ${{\hat{p}}}^{\left( cl \right)}$, it contains two key elements $u$ and ${{\hat{w}}^{\left( cl \right)}}$.
The quantum geometric canonical commutation relation is given by
$\left[ x,{{{\hat{p}}}^{\left( cl \right)}}\left( t \right)\right]=\sqrt{-1}\hbar \left( 1+xu \right)$. As it expresses, $\left[ x,{{{\hat{p}}}^{\left( cl \right)}}\left( t \right)\right]=0$ holds if and only if $u=-x^{-1}, x\neq 0$.

Conclusively, the Schr\"{o}dinger equation and two various kinds of the geometric wave equations are organized as follows:
$\sqrt{-1}\hbar {{\partial }_{t}}\varphi ={{\hat{H}}^{\left( cl \right)}}\varphi=-\frac{{{\hbar }^{2}}}{2m}\varphi_{xx}+\varphi V\left( x \right)$  
and two geometric wave equations 
$\sqrt{-1}\hbar {{\hat{w}}^{\left( cl \right)}}\psi ={{\hat{H}}^{\left( \operatorname{clm} \right)}}\psi =\frac{{{\hbar }^{2}}}{2m}\left( 2u{{\psi }_{x}}+\psi {{u}_{x}} \right)$ and
$ \sqrt{-1}\hbar {{{\hat{w}}}^{\left( ri \right)}}\psi ={{{\hat{H}}}^{\left( \operatorname{rim} \right)}}\psi =\frac{{{\hbar }^{2}}}{2m}\left( 2u{{\psi }_{x}}+\psi {{u}_{x}}+2\psi{{u}^{2}} \right)$ respectively for a better comparison.
Obviously, these three equations can be subtracted in pairs respectively
 \[\sqrt{-1}\hbar \left( {{\partial }_{t}}-{{{\hat{w}}}^{\left( cl\right)}} \right)\psi  =-\frac{{{\hbar }^{2}}}{2m}\left( {{\psi }_{xx}}+2u{{\psi }_{x}}+\psi {{u}_{x}} \right)+\psi V\left( x \right)\]
Similarly,
$\sqrt{-1}\hbar \left( {{\partial }_{t}}-{{{\hat{w}}}^{\left( ri \right)}} \right)\psi =-\frac{{{\hbar }^{2}}}{2m}\left( {{\psi }_{xx}}+2u{{\psi }_{x}}+\psi {{u}_{x}}+2\psi {{u}^{2}} \right)+\psi V\left( x \right)$.
Analogically, for the sum of one pair, two cases emerge, we also accordingly get
$$\sqrt{-1}\hbar \left( {{\partial }_{t}}+{{{\hat{w}}}^{\left( cl \right)}} \right)\psi =-\frac{{{\hbar }^{2}}}{2m}\left({{\psi }_{xx}}- 2u{{\psi }_{x}}-\psi {{u}_{x}}\right)+\psi V\left( x \right)$$ And
$\sqrt{-1}\hbar \left( {{\partial }_{t}}+{{{\hat{w}}}^{\left( ri \right)}} \right)\psi =-\frac{{{\hbar }^{2}}}{2m}\left( {{\psi }_{xx}}-2u{{\psi }_{x}}-\psi {{u}_{x}}-2\psi {{u}^{2}} \right)+\psi V\left( x \right) $.

The positive definite inner product on a Hilbert space [E-P]  for any two states $\phi$ and $\varphi$ is defined as
\begin{equation}\label{a8}
  \left( \phi ,\varphi  \right)=\int{{{\phi }^{*}}\varphi dx}={{\left( \int{\phi {{\varphi }^{*}}dx} \right)}^{*}}={{\left( \varphi ,\phi  \right)}^{*}}
\end{equation}
with the properties as follows:
\begin{description}
  \item[1)]$\left( \phi ,\phi  \right)\ge 0$
  \item[2)] ${{\left( \varphi ,\phi  \right)}^{*}}=\left( \phi ,\varphi  \right)$
  \item[3)] $\left( \phi ,{{c}_{1}}{{\varphi }_{1}}+{{c}_{2}}{{\varphi }_{2}} \right)={{c}_{1}}\left( \phi ,{{\varphi }_{1}} \right)+{{c}_{2}}\left( \phi ,{{\varphi }_{2}} \right)$
  \item[4)] $\left( {{c}_{1}}{{\phi }_{1}}+{{c}_{2}}{{\phi }_{2}},\varphi  \right)=c_{1}^{*}\left( {{\phi }_{1}},\varphi  \right)+c_{2}^{*}\left( {{\phi }_{2}},\varphi  \right)$
\end{description}where ${{c}_{1}},{{c}_{2}}$ are complex constants.  Especially, for the last two properties can be simplified as  $\left( \phi ,{{c}_{1}}\varphi  \right)={{c}_{1}}\left( \phi ,\varphi  \right)$, and ~$\left( {{c}_{1}}\phi ,\varphi  \right)=c_{1}^{*}\left( \phi ,\varphi  \right)$.
Meanwhile, the transpose operator of $\hat{A}$ is given by
\[\left( \phi ,\tilde{\hat{A}}\varphi  \right)=\int{{{\phi }^{*}}\tilde{\hat{A}}\varphi dx}=\left( {{\varphi }^{*}},\hat{A}{{\phi }^{*}} \right)=\int{\varphi \hat{A}{{\phi }^{*}}dx}\]
where the notation $\widetilde{\hat{A}}$ is to denote the transpose of an operator $\hat{A}$.
The adjoint of an operator $\hat{A}$  may also be called the Hermitian conjugate, Hermitian transpose of $\hat{A}$ is denoted by ${{\hat{A}}^{\dagger }}$, where ${\hat{A} }^{*}$ is used to represent the conjugate of $\hat{A}$. First let us define the Hermitian conjugate of an operator $\hat{A}$  to be ${{\hat{A}}^{\dagger }}$.
\[\left( \phi ,{{{\hat{A}}}^{\dagger }}\varphi  \right)=\int{{{\phi }^{*}}{{{\hat{A}}}^{\dagger }}\varphi dx}=\int{\varphi {{\left( \hat{A}\phi  \right)}^{*}}dx}=\left( \hat{A}\phi ,\varphi  \right)\]
That is
\[\left( \phi ,{{{\hat{A}}}^{\dagger }}\varphi  \right)=\left( \hat{A}\phi ,\varphi  \right)={{\left( \varphi ,\hat{A}\phi  \right)}^{*}}=\left( {{\varphi }^{*}},{{{\hat{A}}}^{*}}{{\phi }^{*}} \right)=\left( \phi ,{{\widetilde{{\hat{A}}}}^{*}}\varphi  \right)\]Hence, it deduce a result
${{\hat{A}}^{\dagger }}={{\widetilde{{\hat{A}}}}^{*}}$.  Meanwhile, it has some properties listed as follows:
\begin{description}
  \item[a)] ${{\left( {{{\hat{A}}}^{\dagger }} \right)}^{\dagger }}=\hat{A} $
  \item[b)] ${{\left( {{c}_{1}}\hat{A} \right)}^{\dagger }}={{c}_{1}}^{*}{{{\hat{A}}}^{\dagger }}$
  \item[c)] ${{\left( \hat{A}+\hat{C} \right)}^{\dagger }}={{{\hat{A}}}^{\dagger }}+{{{\hat{C}}}^{\dagger }}$
  \item[d)] ${{\left( \hat{A}\hat{C} \right)}^{\dagger }}={{{\hat{C}}}^{\dagger }}{{{\hat{A}}}^{\dagger }}$
\end{description}
where ${{c}_{1}}^{*}$ denotes the complex conjugate of the complex number ${c}_{1}$.
An Hermitian operator is also call a self-adjoint operator which is a linear operator on a Hilbert space that is equipped with positive definite inner product \eqref{a8},
one can show that for a Hermitian operator in one-dimensional case,
\begin{equation}\label{a3}
  \left( \varphi ,\hat{A}\psi  \right)=\int{{{\varphi }^{*}}\hat{A}}\psi dx=\int{\psi {{\left( \hat{A}\varphi  \right)}^{*}}}dx=\left( \hat{A}\varphi ,\psi  \right)
\end{equation}	
for any two states $\psi$ and $\varphi$. Operators $\hat{A}$ which satisfy this condition are called Hermitian [A-G].  An operator is called Hermitian if $\left( \varphi ,\hat{A}\psi  \right)=\left( \hat{A}\varphi ,\psi  \right)$ holds in a inner product.   In one word, an operator $\hat{O}$ is Hermitian if it is equal to its own conjugate transpose, that is, if
 $\hat{O}^\dagger =\hat{O}$, then $\hat{O}$ is Hermitian denoted as $\hat{O}\in Her$. The important properties of Hermitian operators include: [B-L],[V-M],[N-G]\par
1). real eigenvalues,\par
2). eigenfunctions with different eigenvalues are orthogonal, \par
3). eigenfunctions can be chosen to be a complete orthonormal basis,\par
In some sense, these Hermitian operators play the role of the real numbers (being equal to their own complex conjugate) and form a real vector space. They serve as the model of real-valued observables in quantum mechanics.
Actual measurements are directly related to Hermitian operators. Indeed, Hermitian operators have only real eigenvalues. Meanwhile, in contrast to Hermitian operators,  an operator $\hat{B}$ is said to be a skew-Hermitian operator if it satisfies
\[\left( \phi ,\hat{B}\varphi  \right)=\int{{{\phi }^{*}}\hat{B}\varphi dx}=\int{\varphi {{\left( -\hat{B}\phi  \right)}^{*}}dx}=-\int{\varphi {{\left( \hat{B}\phi  \right)}^{*}}dx}\]Namely $\left( \phi ,\hat{B}\varphi  \right)=-\left( \hat{B}\phi ,\varphi  \right)$ holds.
As mentioned previously, the expectation value of an operator $\hat{A}$ is given by
$\left\langle A \right\rangle =\int{{{\psi }^{*}}\hat{A}}\psi dx=\left( \psi ,\hat{A}\psi  \right)$,
and all physical observables are represented by such expectation values. Obviously, the value of a physical observable such as energy or density must be real, so we require $\left\langle A \right\rangle$ to be real.
In many applications, we are led to consider operators that are unbounded; examples include the position, momentum, and Hamiltonian operators in quantum mechanics, as well as many differential operators. In the unbounded case, there are a number of subtle technical issues that have to be dealt with. In particular, there is a crucial distinction between operators that are merely symmetric and those that are self-adjoint.

Judging the Hermitian property of quantum operators is very important for the study of quantum mechanics, especially for the study of eigenvalue equations, which can help us determine the properties of eigenvalues and the relevant information of eigenfunctions.

Using the definition of the Hermitian operator in one-dimensional case,  we obtain the following results.
\begin{theorem}\label{t1}
 The G-dynamics \eqref{a1} ${{\hat{w}}^{\left( cl \right)}}\in Her$ is a Hermitian operator, then its eigenvalues ${{w}^{\left( q \right)}}$ are real.
\end{theorem}
Note that the quantum physical observables in the experiment are Hermitian operators,  once we prove the G-dynamics \eqref{a1} is a Hermitian operator, as a result, it's a quantum physical observables in the experiment,  then it will help us better understand how it can be observed, because it's a observable. 
As a consequence of ${{\hat{w}}^{\left( cl \right)}}\in Her$, then it yields  
${{\hat{w}}^{\left( cl \right)}}={{\hat{w}}^{\left( cl \right)\dagger }}\in Her$ and 
\begin{align}
  & {{\left( {{c}_{1}}{{{\hat{w}}}^{\left( cl \right)}} \right)}^{\dagger }}={{c}_{1}}^{*}{{{\hat{w}}}^{\left( cl \right)}} \notag\\ 
 & {{\left( {{{\hat{w}}}^{\left( cl \right)\dagger }} \right)}^{\dagger }}={{{\hat{w}}}^{\left( cl \right)}}\in Her \notag 
\end{align}and $\left( {{{\hat{w}}}^{\left( cl \right)}}\phi ,\varphi  \right)=\left( \phi ,{{{\hat{w}}}^{\left( cl \right)}}\varphi  \right)$ always holds  for any two states $\phi$ and $\varphi$.   
It's well-known that the observables correspond to self-adjoint operators which are used in functional analysis and quantum mechanics.
\begin{corollary}\label{c1}
 Operator ${{\hat{w}}^{\left( ri \right)}}={{\hat{w}}^{\left( cl \right)}}+{{w}^{\left( s \right)}}\in NHer$ is a non-Hermitian operator, then the eigenvalues are in a complex form.
\end{corollary}Note that non-locality is as common as non-Hermiticity, it has a necessary to study how it represents in quantum mechanics.
Skew-Hermitian operator can be understood as the complex versions of real skew-symmetric operator. The eigenvalues of a skew-Hermitian operator are all purely imaginary (and possibly zero). An arbitrary operator $\hat{A}\in NHer$ can be written as the sum of a Hermitian operator $\hat{C}\in Her$  and a skew-Hermitian operator $\hat{B}\in SHer$, that is to say,
$\hat{A}=\hat{C}+\hat{B}$, where
$\hat{C}=\left( \hat{A}+{{\hat{A}}^{\dagger }} \right)/2,~~\hat{B}=\left( \hat{A}-{{\hat{A}}^{\dagger }} \right)/2$.  In general, the skew-Hermitian operator $\hat{B}\neq 0$ which means $\hat{A}\neq{{\hat{A}}^{\dagger }} $,

\begin{theorem}\label{t2}
For the G-dynamics \eqref{a1}, the following statements are equivalent:
\begin{enumerate}
    \item ${{\hat{w}}^{\left( cl \right)}}\in Her$,
    \item ${{w}^{\left( q \right)}}$ are real.
    \item $\sqrt{-1}{{\hat{w}}^{\left( cl \right)}}\in SHer$ if and only if ${{\hat{w}}^{\left( cl \right)}}\in Her$.
    \item ${{\hat{w}}^{\left( cl \right)}}\in Her$ if and only if $\sqrt{-1}{{\hat{w}}^{\left( cl \right)}}\in SHer$.

  \end{enumerate}
\end{theorem}
Notice that ${{\hat{w}}^{\left( cl \right)}}$ is a Hermitian operator satisfying $\left( {{{\hat{w}}}^{\left( cl \right)}}\varphi ,\psi  \right)=\left( \varphi ,{{{\hat{w}}}^{\left( cl \right)}}\psi  \right)$, the operator $\hat{B}\in SHer $ if it satisfies the relation ${{\hat{B}}^{\dagger }}=-\hat{B}$, more precisely, it satisfies the relation $\left( \varphi ,\hat{B}\psi  \right)=-\left( \hat{B}\varphi ,\psi  \right)$,  as a result of this definition, therefore,  $\sqrt{-1}{{\hat{w}}^{\left( cl \right)}}$ is skew-Hermitian fulfilling $$\left( \sqrt{-1}{{{\hat{w}}}^{\left( cl \right)}}\varphi ,\psi  \right)=-\left( \varphi ,\sqrt{-1}{{{\hat{w}}}^{\left( cl \right)}}\psi  \right)$$ for any two states $\psi$ and $\varphi$.

\begin{theorem}
  Let ${{\hat{w}}^{\left( cl \right)}}\in Her$ be given, then the eigenvalues of
${{\hat{H}}^{\left( \operatorname{clm} \right)}}=\sqrt{-1}\hbar {{\hat{w}}^{\left( cl \right)}}\in SHer$ are all purely imaginary.
\end{theorem}
Note that as we can see the Hermitian proof of the ${{\hat{w}}^{\left( cl \right)}}\in Her$ is the key to verify the real or the complex of its eigenvalues.
\begin{corollary}
 Based on ${{\hat{w}}^{\left( ri \right)}}\in NHer$, then eigenvalues of
${{\hat{H}}^{\left( \operatorname{rim} \right)}}\in NHer$ are all complex form.
\end{corollary}
If an operator is Hermitian, then the eigenvalues are real. In bound systems, the eigenvalues are discrete.
\begin{theorem}\label{t3}
 The curvature operator $\hat{Q}\in SHer$ is a skew-Hermitian operator, then the eigenvalues of it are all purely imaginary.
\end{theorem}
Note that the theorem \ref{t1} reflects the theorem \ref{t3}, and vice versa. Such property of the curvature operator $\hat{Q}\in SHer$ actually captures our attentions, it's weird and interesting property.
\begin{theorem}\label{t5}
The one-dimensional motor operator \eqref{a7} ${{\hat{E}}^{\left( w \right)}}=-\frac{{{\hbar }^{2}}}{2m}{d}^{2}/d{{x}^{2}}-{{\hat{H}}^{\left( \operatorname{clm} \right)}}\in NHer$ holds, where $-\frac{{{\hbar }^{2}}}{2m}\frac{{{d}^{2}}}{d{{x}^{2}}}\in Her$ and $ {{\hat{H}}^{\left( \operatorname{clm} \right)}}\in SHer$, then the eigenvalues are in a complex form.
\end{theorem}

\begin{corollary} The  \eqref{a2}  ${{\hat{H}}^{\left( ri \right)}} ={{\hat{H}}^{\left( g \right)}} -{{\hat{H}}^{\left( \operatorname{clm} \right)}}\in NHer$ always holds, where ${{\hat{H}}^{\left( g \right)}} \in Her$ and
  ${{\hat{H}}^{\left( \operatorname{clm} \right)}}\in SHer$.

\end{corollary}
Note that there is an axiom of quantum mechanics requiring that Hamiltonian is the
Hermitian due to the Hermiticity guarantees that the energy spectrum remains real, but the Hermiticity are badly restricted to study the non-linear quantum phenomena.

\section{Proofs of the main results}

\par
 In this section we prove the statement of theorem \ref{t1}. To start with a lemma will be used in the proof of the theorem \ref{t1}.
\begin{lemma}\label{l1}
 For a given function $f$, then ${{\hat{w}}^{\left( cl \right)}}f={{b}_{c}}\left( 2d\left( uf \right)/dx-{{u}_{x}}f \right)$.\par
  \begin{proof}
 By derivative calculus, it directly evaluates
  \begin{align}
  {{{\hat{w}}}^{\left( cl \right)}}f& ={{b}_{c}}\left( 2udf/dx+f{{u}_{x}} \right)={{b}_{c}}\left( 2\left( d\left( fu \right)/dx-f{{u}_{x}} \right)+f{{u}_{x}} \right)  \notag\\
 & ={{b}_{c}}\left( 2d\left( uf \right)/dx-{{u}_{x}}f \right) \notag
\end{align}
Therefore, the simple proof is completed.

  \end{proof}

\end{lemma}
Note that square integrability requires that the wave function tends to 0 at infinity. This will be used through the proof below.

\subsection{Proof of the theorem \ref{t1}}
\par
{\bf Proof of the theorem \ref{t1}}:
\begin{proof}
Based on the definition of the Hermitian operator \eqref{a3} in one-dimensional case, it just derives a result as follows,
\begin{align}
 \left( \varphi ,{{{\hat{w}}}^{\left( cl \right)}}\psi  \right) &=\int{{{\varphi }^{*}}{{{\hat{w}}}^{\left( cl \right)}}\psi dx} =\int{{{\varphi }^{*}}\left( {{b}_{c}}\left( 2u{{\psi }_{x}}+\psi {{u}_{x}} \right) \right)dx} \notag\\
 &=2{{b}_{c}}\int{{{\varphi }^{*}}u{{\psi }_{x}}dx}+{{b}_{c}}\int{{{\varphi }^{*}}{{u}_{x}}\psi dx}  \notag\\
 & =-2{{b}_{c}}\int{\psi u{{\varphi }_{x}}^{*}}dx-2{{b}_{c}}\int{\psi {{u}_{x}}{{\varphi }^{*}}}dx+{{b}_{c}}\int{{{\varphi }^{*}}{{u}_{x}}\psi dx}  \notag
\end{align}
where we have used the lemma \ref{l1}
\begin{align}
 \int{{{\varphi }^{*}}u{{\psi }_{x}}dx} &=\int{{{\varphi }^{*}}ud\psi }=u{{\varphi }^{*}}\psi -\int{\psi d\left( {{\varphi }^{*}}u \right)}=-\int{\psi d\left( {{\varphi }^{*}}u \right)}  \notag\\
 & =-\int{\psi \left( {{\varphi }_{x}}^{*}u+{{\varphi }^{*}}{{u}_{x}} \right)}dx  \notag\\
 & =-\int{\psi u{{\varphi }_{x}}^{*}}dx-\int{\psi {{u}_{x}}{{\varphi }^{*}}}dx  \notag
\end{align}
for above derivations, and we have used integration by parts, of course, and threw away the boundary term for the usual reason: if $\varphi$ and $\psi$ are square integrable, they must go to zero at $\pm \infty$, that is to say, ${{\varphi }^{*}}\psi \left| _{\infty } \right.=0$,  hence, then it further leads to
\begin{align}
 \int{{{\varphi }^{*}}{{{\hat{w}}}^{\left( cl \right)}}\psi dx}& =-2{{b}_{c}}\int{\psi u{{\varphi }_{x}}^{*}}dx-{{b}_{c}}\int{\psi {{u}_{x}}{{\varphi }^{*}}dx}  \notag\\
 & =\int{\psi \left( -2{{b}_{c}}u \right){{\varphi }_{x}}^{*}}dx+\int{\psi \left( -{{b}_{c}}{{u}_{x}} \right){{\varphi }^{*}}dx}  \notag\\
 & =\int{\psi {{\left( 2{{b}_{c}}u{{\varphi }_{x}} \right)}^{*}}}dx+\int{\psi {{\left( {{b}_{c}}{{u}_{x}}\varphi  \right)}^{*}}dx}  \notag\\
 & =\int{\psi {{\left( 2{{b}_{c}}u{{\varphi }_{x}}+{{b}_{c}}{{u}_{x}}\varphi  \right)}^{*}}}dx  \notag\\
 & =\int{\psi {{\left( {{{\hat{w}}}^{\left( cl \right)}}\varphi  \right)}^{*}}}dx =\left( {{{\hat{w}}}^{\left( cl \right)}}\varphi ,\psi  \right) \notag
\end{align}As a consequence of the Hermitian condition $\left( {{{\hat{w}}}^{\left( cl \right)}}\varphi ,\psi  \right)=\left( \varphi ,{{{\hat{w}}}^{\left( cl \right)}}\psi  \right)$ in terms of the G-dynamics \eqref{a1}, then the G-dynamics \eqref{a1} is a Hermitian operator, namely,  ${{\hat{w}}^{\left( cl \right)}}={{\hat{w}}^{\left( cl \right)\dagger }}\in Her$, in accordance with the property of the Hermitian operator, then its eigenvalues ${{w}^{\left( q \right)}}$ are real.  Hence, we complete the proof.
\end{proof}
Without loss of generality, we consider the general case of the Hermiticity, that is, the case of the non-Hermitian operators.
\begin{theorem}\label{t4}
If $\hat{a}\in NHer$ holds, then it can be decomposed into complex form
$\hat{a}={{\hat{a}}^{\left( h \right)}}+{{\hat{a}}^{\left( sh \right)}}\in NHer$, where  ${{\hat{a}}^{\left( sh \right)}}=\sqrt{-1}{{\hat{a}}^{\left( hi \right)}}\in SHer$, and
\begin{equation}\label{a6}
  {{\hat{a}}^{\left( h \right)}}=\frac{1}{2}\left( \hat{a}+{{\hat{a}}^{\dagger }} \right)\in Her,~~{{\hat{a}}^{\left( hi \right)}}=\frac{1}{2\sqrt{-1}}\left( \hat{a}-{{\hat{a}}^{\dagger }} \right)\in Her
\end{equation}
 \begin{proof}
  As the non-Hermitian operators defined, it has the form $\hat{a}={{\hat{a}}^{\left( h \right)}}+{{\hat{a}}^{\left( sh \right)}}={{\hat{a}}^{\left( h \right)}}+\sqrt{-1}{{\hat{a}}^{\left( hi \right)}}\in NHer$, where ${{\hat{a}}^{\left( h \right)}}={{\hat{a}}^{\left( h\right)\dagger}}$,~~${{\hat{a}}^{\left( sh \right)}}=-{{\hat{a}}^{\left( sh\right)\dagger}}$, then it gets
  ${{\hat{a}}^{\dagger }}={{\hat{a}}^{\left( h \right)}}-\sqrt{-1}{{\hat{a}}^{\left( hi \right)}}$.  As a result, it leads to the Hermitian operators
  ${{\hat{a}}^{\left( h \right)}}=\frac{1}{2}\left( \hat{a}+{{\hat{a}}^{\dagger }} \right)\in Her$ and ${{\hat{a}}^{\left( hi \right)}}=\frac{1}{2\sqrt{-1}}\left( \hat{a}-{{\hat{a}}^{\dagger }} \right)\in Her$.  For these Hermitian operators, it can be easily verified, that is
${{\hat{a}}^{\left( h \right)\dagger }}={{\hat{a}}^{\left( h \right)}}$ and
$ {{\hat{a}}^{\left( hi \right)\dagger }}={{\hat{a}}^{\left( hi \right)}}$.
Hence, we complete the proof.
\end{proof}
\end{theorem}
Obviously, the non-Hermitian operator must fulfill ${{\hat{a}}^{\left( sh \right)}}\neq 0$, or ${{\hat{a}}^{\left( hi \right)}}\neq 0$, that is to say, $\hat{a}\neq{{\hat{a}}^{\dagger }}$, it can be seen as skew Hermitian operator ${{\hat{a}}^{\left( sh \right)}}=\sqrt{-1}{{\hat{a}}^{\left( hi \right)}}\in SHer$ breaks the symmetry of Hermitian operator ${{\hat{a}}^{\left( h \right)}}\in Her$.
In particular, based on ${{\hat{a}}^{\left( sh \right)}}=\sqrt{-1}{{\hat{a}}^{\left( hi \right)}}\in SHer$, we think of its converse case, that is to say,
${{\hat{a}}^{\left( hi \right)}}=-\sqrt{-1}{{\hat{a}}^{\left( sh \right)}}\in Her$, and for the non-Hermitian operator
$\hat{a}={{\hat{a}}^{\left( h \right)}}+\sqrt{-1}{{\hat{a}}^{\left( hi \right)}}\in NHer$, we also have the same operation given by
$\sqrt{-1}\hat{a}=\sqrt{-1}{{\hat{a}}^{\left( h \right)}}-{{\hat{a}}^{\left( hi \right)}}$, and $\sqrt{-1}{{\hat{a}}^{\left( h \right)}}\in SHer$ follows.
Hence, we can say that if $\hat{A}\in Her$, then $\sqrt{-1}\hat{A}\in SHer$;  and  if $\hat{A}\in SHer$, then $\sqrt{-1}\hat{A}\in Her$,   and vice versa.
Taking the non-Hermitian operator $\hat{a}$ into consideration by using definition \eqref{a3} leads to the result
\begin{align}
 \left( \varphi ,\hat{a}\psi  \right) &=\int{{{\varphi }^{*}}\hat{a}\psi dx}=\int{{{\varphi }^{*}}\left( {{\hat{a}}^{\left( h \right)}}+{{\hat{a}}^{\left( sh \right)}} \right)\psi dx}=\left( \varphi ,{{\hat{a}}^{\left( h \right)}}\psi +{{\hat{a}}^{\left( sh \right)}}\psi  \right)\notag\\
 &=\left( \varphi ,{{\hat{a}}^{\left( h \right)}}\psi  \right)+\left( \varphi ,{{\hat{a}}^{\left( sh \right)}}\psi  \right)=\left( \varphi ,{{\hat{a}}^{\left( h \right)}}\psi  \right)+\left( \varphi ,\sqrt{-1}{{\hat{a}}^{\left( hi \right)}}\psi  \right)\notag\\
 & =\left( {{\hat{a}}^{\left( h \right)}}\varphi ,\psi  \right)-\left( \sqrt{-1}{{\hat{a}}^{\left( hi \right)}}\varphi ,\psi  \right) \notag\\
 & =\left( \left( {{\hat{a}}^{\left( h \right)}}-\sqrt{-1}{{\hat{a}}^{\left( hi \right)}} \right)\varphi ,\psi  \right) \notag\\
 & =\left( {{\hat{a}}^{\dagger }}\varphi ,\psi  \right)\notag
\end{align}
where $$\left( \varphi ,\sqrt{-1}{{\hat{a}}^{\left( hi \right)}}\psi  \right)=\sqrt{-1}\left( \varphi ,{{\hat{a}}^{\left( hi \right)}}\psi  \right)=\sqrt{-1}\left( {{\hat{a}}^{\left( hi \right)}}\varphi ,\psi  \right)=-\left( \sqrt{-1}{{\hat{a}}^{\left( hi \right)}}\varphi ,\psi  \right)$$ and    ${{\hat{a}}^{\dagger }}={{\hat{a}}^{\left( h \right)}}-\sqrt{-1}{{\hat{a}}^{\left( hi \right)}}\in NHer$ is the conjugate transposes of the non-Hermitian operator $\hat{a}\in NHer$. Therefore, we can verify the Hermiticity of ${{{\hat{H}}}^{\left( ri \right)}}$ based on the \eqref{a3},
\begin{align}
 \left( \varphi ,{{{\hat{H}}}^{\left( ri \right)}}\psi  \right) & =\int{{{\varphi }^{*}}{{{\hat{H}}}^{\left( ri \right)}}\psi dx}=\int{{{\varphi }^{*}}\left( {{{\hat{H}}}^{\left( cl \right)}}-{{E}^{\left( s \right)}}/2-\sqrt{-1}\hbar {{{\hat{w}}}^{\left( cl \right)}} \right)\psi dx}  \notag\\
 & =\int{{{\varphi }^{*}}{{{\hat{H}}}^{\left( cl \right)}}\psi dx}-\frac{1}{2}\int{{{\varphi }^{*}}{{E}^{\left( s \right)}}\psi dx}-\sqrt{-1}\hbar \int{{{\varphi }^{*}}{{{\hat{w}}}^{\left( cl \right)}}\psi dx}  \notag\\
 & =\left( \varphi ,{{{\hat{H}}}^{\left( cl \right)}}\psi  \right)-\frac{1}{2}\left( \varphi ,{{E}^{\left( s \right)}}\psi  \right)-\sqrt{-1}\hbar \left( \varphi ,{{{\hat{w}}}^{\left( cl \right)}}\psi  \right)  \notag\\
 & =\left( {{{\hat{H}}}^{\left( cl \right)}}\varphi ,\psi  \right)-\frac{1}{2}\sqrt{-1}\hbar \left( \varphi ,{{w}^{\left( s \right)}}\psi  \right)+\left( \sqrt{-1}\hbar {{{\hat{w}}}^{\left( cl \right)}}\varphi ,\psi  \right)  \notag\\
 & =\left( {{{\hat{H}}}^{\left( cl \right)}}\varphi ,\psi  \right)+\frac{1}{2}\left( -\sqrt{-1}\hbar {{w}^{\left( s \right)}}\varphi ,\psi  \right)+\left( \sqrt{-1}\hbar {{{\hat{w}}}^{\left( cl \right)}}\varphi ,\psi  \right)  \notag\\
 & =\left( {{{\hat{H}}}^{\left( cl \right)}}\varphi ,\psi  \right)+\left( -{{E}^{\left( s \right)}}\varphi /2,\psi  \right)+\left( \sqrt{-1}\hbar {{{\hat{w}}}^{\left( cl \right)}}\varphi ,\psi  \right)  \notag\\
 & =\left( \left( {{{\hat{H}}}^{\left( cl \right)}}-{{E}^{\left( s \right)}}/2+\sqrt{-1}\hbar {{{\hat{w}}}^{\left( cl \right)}} \right)\varphi ,\psi  \right)  \notag\\
 & =\left( {{{\hat{H}}}^{\left( ri \right)\dagger }}\varphi ,\psi  \right) \notag
\end{align}
where ${{E}^{\left( s \right)}}=\sqrt{-1}\hbar {{w}^{\left( s\right)}}=2\sqrt{-1}\hbar {{b}_{c}}{{u}^{2}}\in Her$, and
$\left( {{w}^{\left( s \right)}}\varphi ,\psi  \right)=-\left( \varphi ,{{w}^{\left(s \right)}}\psi\right)$, and the properties of the positive definite inner product have been used. For the precise analysis of the non-Hermitian Hamiltonian operator \eqref{a2}, it shows 
\[\underbrace{{{{\hat{H}}}^{\left( ri \right)}}}_{\text{non-Hermitian Hamiltonian}}=\underbrace{\underbrace{\underbrace{{{{\hat{H}}}^{\left( cl \right)}}}_{\text{ classical Hamiltonian operator}}-\underbrace{\frac{{{E}^{\left( s \right)}}}{2}}_{\text{ line curvature}~u}}_{={{{\hat{H}}}^{\left( g \right)}}\text{ is Hermitian}}-\underbrace{\underbrace{\sqrt{-1}\hbar {{{\hat{w}}}^{\left( cl \right)}}}_{\text{ G-dynamics }{{{\hat{w}}}^{\left( cl \right)}}}}_{\text{ skew-Hermitian}}}_{\text{non-Hermitian}}\]
As an application, we put the theorem \ref{t4} into practice, by considering the conjugate transposes of the non-Hermitian Hamiltonian operator \eqref{a2} ${{\hat{H}}^{\left( ri \right)}}\in NHer$ that is given by
\begin{align}
  {{{\hat{H}}}^{\left( ri \right)\dagger }}& ={{{\hat{H}}}^{\left( cl \right)\dagger }}-{{E}^{\left( s \right)\dagger }}/2-{{\left( \sqrt{-1}\hbar {{{\hat{w}}}^{\left( cl \right)}} \right)}^{\dagger }}={{{\hat{H}}}^{\left( cl \right)}}-{{E}^{\left( s \right)}}/2+\sqrt{-1}\hbar {{{\hat{w}}}^{\left( cl \right)\dagger }}  \notag\\
 & ={{\hat{H}}^{\left( g \right)}} +\sqrt{-1}\hbar {{{\hat{w}}}^{\left( cl \right)}} \notag
\end{align}where we have used ${{\hat{H}}^{\left( cl \right)}}, {{\hat{H}}^{\left( g \right)}} \in Her$.
Then it generally results in an outcome by an operation of difference   \[0\neq\Delta {{\hat{H}}^{\left( ri \right)}}={{\hat{H}}^{\left( ri \right)}}-{{\hat{H}}^{\left( ri \right)\dagger }}=-2\sqrt{-1}\hbar {{\hat{w}}^{\left( cl \right)}}=-2{{\hat{H}}^{\left( \operatorname{clm} \right)}}\in SHer\]
Conversely, $-\Delta {{\hat{H}}^{\left( ri \right)}}=2\sqrt{-1}\hbar {{\hat{w}}^{\left( cl \right)}}=2{{\hat{H}}^{\left( \operatorname{clm} \right)}}$, based on \eqref{a6}, it nicely conforms to self consistency $\frac{\Delta {{{\hat{H}}}^{\left( ri \right)}}}{2\sqrt{-1}}=-\hbar {{\hat{w}}^{\left( cl \right)}}\in Her$,  furthermore, it forms the Hermitian operators  \[\sqrt{-1}\Delta {{\hat{H}}^{\left( ri \right)}}/2=\sqrt{-1}\left( {{{\hat{H}}}^{\left( ri \right)}}-{{{\hat{H}}}^{\left( ri \right)\dagger }} \right)/2=\hbar {{\hat{w}}^{\left( cl \right)}}\in Her\]as technical operation derives, meanwhile, it yields skew Hermitian operator
$\frac{\Delta {{{\hat{H}}}^{\left( ri \right)}}}{2\hbar }=-\sqrt{-1} {{\hat{w}}^{\left( cl \right)}}\in SHer$.
We also get the sum of the non-Hermitian Hamiltonian operator \eqref{a2} and its  the conjugate transposes form that is expressed as  $${{\hat{H}}^{\left( ri \right)}}+{{\hat{H}}^{\left( ri \right)\dagger }}=2{{\hat{H}}^{\left( cl \right)}}-{{E}^{\left( s \right)}}=2{{\hat{H}}^{\left( g \right)}}\in Her $$
As a consequence, for any non-Hermitian Hamiltonian operator ${{\hat{H}}^{\left( ri \right)}}$, it can be divided into two parts: the Hermitian operator and the skew-Hermitian operator \[{{\hat{H}}^{\left( ri \right)}}=\frac{{{{\hat{H}}}^{\left( ri \right)}}+{{{\hat{H}}}^{\left( ri \right)\dagger }}}{2}+\frac{\Delta {{{\hat{H}}}^{\left( ri \right)}}}{2}={{\hat{H}}^{\left( g \right)}} -\sqrt{-1}\hbar {{\hat{w}}^{\left( cl \right)}}\in NHer\]
Clearly, we can verify the part of the Hermitian operator \[{{\left( {{{\hat{H}}}^{\left( ri \right)}}+{{{\hat{H}}}^{\left( ri \right)\dagger }} \right)}^{\dagger }}/2={{\left( {{{\hat{H}}}^{\left( cl \right)}}-{{E}^{\left( s \right)}}/2 \right)}^{\dagger }}={{\hat{H}}^{\left( cl \right)}}-{{E}^{\left( s \right)}}/2={{\hat{H}}^{\left( g \right)}} \in Her\]while the skew-Hermitian operator satisfies
\begin{equation}\label{a5}
  {{\left( \Delta {{{\hat{H}}}^{\left( ri \right)}} \right)}^{\dagger }}/2={{\left( -\sqrt{-1}\hbar {{{\hat{w}}}^{\left( cl \right)}} \right)}^{\dagger }}=\sqrt{-1}\hbar {{\hat{w}}^{\left( cl \right)\dagger }}=\sqrt{-1}\hbar {{\hat{w}}^{\left( cl \right)}}\in SHer
\end{equation}
As the theorem \ref{t4} proves, we can conclude that the non-Hermitian Hamiltonian operator ${{\hat{H}}^{\left( ri \right)}}$  \eqref{a2} is a complete operator, and the main part is the one-dimensional motor operator ${{\hat{E}}^{\left( w \right)}}$ as the derivative operator while the rest remains as the function.
Specially, the theorem \ref{t4} implies that it requires the necessity of the skew-Hermitian operator ${{\hat{H}}^{\left( \operatorname{clm} \right)}}=\sqrt{-1}\hbar {{\hat{w}}^{\left( cl \right)}}\neq 0\in SHer$ as a quantum correction term to make quantum theory more complete, it depicts the quantum phenomena such as the quantum rotation-like about the quantum space.  As a result, the non-Hermitian Hamiltonian operator ${{\hat{H}}^{\left( ri \right)}}$  \eqref{a2} deduces the corresponding the geometric frequency operator
\begin{align}
 {{\hat{f}}^{\left( ri \right)}} & ={{{\hat{H}}}^{\left( ri \right)}}/\hbar ={{{\hat{H}}}^{\left( cl \right)}}/\hbar -\frac{{{E}^{\left( s \right)}}}{2\hbar }-\sqrt{-1}{{{\hat{w}}}^{\left( cl \right)}} \notag\\
 & ={{{\hat{H}}}^{\left( cl \right)}}/\hbar -\sqrt{-1}{{w}^{\left( s \right)}}/2-\sqrt{-1}{{{\hat{w}}}^{\left( cl \right)}} \notag
\end{align}where ${{\hat{f}}^{\left( ri \right)}}\in NHer$, thusly, the Hermitian operator and skew Hermitian operator in the name of the geometric frequency operator can be given respectively
\begin{align}
  & {{\hat{f}}^{\left( g \right)}}={{{\hat{H}}}^{\left( g \right)}}/\hbar ={{{\hat{H}}}^{\left( cl \right)}}/\hbar -\sqrt{-1}{{w}^{\left( s \right)}}/2\in Her, \notag\\
 & {{{\hat{H}}}^{\left( \operatorname{clm}  \right)}}/\hbar =\sqrt{-1}{{{\hat{w}}}^{\left( cl \right)}}\in SHer \notag
\end{align}Therefore, in a simple word, the geometric frequency operator is compactly rewritten as
$${{\hat{f}}^{\left( ri \right)}}={{\hat{f}}^{\left( g \right)}}-\sqrt{-1}{{\hat{w}}^{\left( cl \right)}}\in NHer$$ and  its Hermitian conjugate  ${{\hat{f}}^{\left( ri \right)\dagger }}={{\hat{f}}^{\left( g \right)}}+\sqrt{-1}{{\hat{w}}^{\left( cl \right)}}\in NHer$ follows, then according to the theorem \ref{t4}, it derives $$0\neq{{\hat{f}}^{\left( ri \right)}}-{{\hat{f}}^{\left( ri \right)\dagger }}=-2\sqrt{-1}{{\hat{w}}^{\left( cl \right)}}\in SHer$$ and based on formula \eqref{a6}, it gets the unique Hermitian expression $\frac{{{{\hat{f}}}^{\left( ri \right)}}-{{{\hat{f}}}^{\left( ri \right)\dagger }}}{2\sqrt{-1}}=-{{\hat{w}}^{\left( cl \right)}}\in Her$. More importantly, conversely, we have 
\[{{\hat{H}}^{\left( ri \right)}}=\hbar {{\hat{f}}^{\left( ri \right)}}=\underbrace{{{{\hat{H}}}^{\left( cl \right)}}}_{\text{ classical Hamiltonian operator}}-\underbrace{\frac{{{E}^{\left( s \right)}}}{2}}_{\text{ line curvature}~u}-\underbrace{\sqrt{-1}\hbar {{{\hat{w}}}^{\left( cl \right)}}}_{\text{ G-dynamics }{{{\hat{w}}}^{\left( cl \right)}}}\]

In the same reason for the motor operator ${{\hat{E}}^{\left( w \right)}}$, it also gets
\[{{\hat{f}}^{\left( w \right)}}={{\hat{E}}^{\left( w \right)}}/\hbar =-\frac{\hbar }{2m}{{d}^{\text{2}}}/d{{x}^{\text{2}}}-\sqrt{-1}{{\hat{w}}^{\left( cl \right)}}\in NHer\]
An important property of Hermitian operators is that their eigenvalues are real.
It's well known that physical observables (e.g., energy, momentum, position, etc.) are represented mathematically by operators in quantum mechanics, For instance, the operator corresponding to energy is the classical Hamiltonian operator.
As a result of the non-Hermitian Hamiltonian operator \eqref{a2}, it implies that its eigenvalues are in a complex form, and the imaginary part is provided by the skew-Hermitian operators \eqref{a5}.

As previously mentioned, all physical observables are represented by such expectation values, and the expectation value of Hermitian operator ${{\hat{w}}^{\left( cl \right)}}$ is given by
$\left\langle {{{w}}^{\left( cl \right)}} \right\rangle =\int{{{\psi }^{*}}{{\hat{w}}^{\left( cl \right)}}}\psi dx=\left( \psi ,{{\hat{w}}^{\left( cl \right)}}\psi  \right)$, more precisely, bring the \eqref{a1} into the expectation value of ${{\hat{w}}^{\left( cl \right)}}\in Her$, then
\begin{align}
 \left\langle {{w}^{\left( cl \right)}} \right\rangle & =\left( \psi ,{{{\hat{w}}}^{\left( cl \right)}}\psi  \right)=\left( \psi ,{{b}_{c}}\left( 2u\frac{d}{dx}+{{u}_{x}} \right)\psi  \right)={{b}_{c}}\left( \psi ,2u{{\psi }_{x}}+\psi {{u}_{x}} \right) \notag\\
 & =2{{b}_{c}}\left( \psi ,u{{\psi }_{x}} \right)+{{b}_{c}}\left( \psi ,{{u}_{x}}\psi  \right) \notag\\
 & =2{{b}_{c}}\left( \psi ,u{{\psi }_{x}} \right)+{{b}_{c}}\left\langle {{u}_{x}} \right\rangle  \notag
\end{align}
where $\left\langle {{u}_{x}} \right\rangle =\left( \psi ,{{u}_{x}}\psi  \right)$ is
the expectation value of ${u}_{x}$.

\vspace{.4 cm}
\par
{\bf Proof of the corollary \ref{c1}}:\par
\begin{proof}
Since the one-dimensional G-dynamics ${{\hat{w}}^{\left( cl \right)}}\in Her$, then ${{\hat{w}}^{\left( cl \right)}}={{\hat{w}}^{\left( cl \right)\dagger }}$, therefore, in accordance with the definition of \eqref{a3}, we get
\begin{align}
 \left( \varphi ,{{{\hat{w}}}^{\left( ri \right)}}\psi  \right) &=\int{{{\varphi }^{*}}{{{\hat{w}}}^{\left( ri \right)}}\psi dx}=\int{{{\varphi }^{*}}\left( {{{\hat{w}}}^{\left( cl \right)}}+{{w}^{\left( s \right)}} \right)\psi dx} \notag\\
 & =\int{{{\varphi }^{*}}{{{\hat{w}}}^{\left( cl \right)}}\psi dx}+\int{{{\varphi }^{*}}{{w}^{\left( s \right)}}\psi dx} \notag\\
 & =\left( \varphi ,{{{\hat{w}}}^{\left( cl \right)}}\psi  \right)+\left( \varphi ,{{w}^{\left( s \right)}}\psi  \right)=\left( {{{\hat{w}}}^{\left( cl \right)}}\varphi ,\psi  \right)+\left( \varphi ,{{w}^{\left( s \right)}}\psi  \right) \notag
\end{align}
due to ${{w}^{\left( s \right)}}=2{{b}_{c}}{{u}^{2}}$ is a pure imaginary function, hence, we firstly verify
\[\left( \varphi ,{{w}^{\left( s \right)}}\psi  \right)=\int{{{\varphi }^{*}}{{w}^{\left( s \right)}}\psi dx}=2{{b}_{c}}\int{{{\varphi }^{*}}{{u}^{2}}\psi dx}=2{{b}_{c}}\left( \varphi ,{{u}^{2}}\psi  \right)\]
and then it's led to the result of skew Hermitian
\begin{align}
 \left( {{w}^{\left( s \right)}}\varphi ,\psi  \right) & =\int{{{\left( {{w}^{\left( s \right)}}\varphi  \right)}^{*}}\psi dx}=\int{{{\left( 2{{b}_{c}}{{u}^{2}}\varphi  \right)}^{*}}\psi dx} \notag\\
 & =-2{{b}_{c}}\int{{{\varphi }^{*}}{{u}^{2}}\psi dx}\notag\\
 &=-\left( \varphi ,{{w}^{\left( s \right)}}\psi  \right) \notag
\end{align}where \[\left( {{w}^{\left( s \right)}}\varphi ,\psi  \right)=2\left( {{b}_{c}}{{u}^{2}}\varphi ,\psi  \right)=-2{{b}_{c}}\int{{{\varphi }^{*}}{{u}^{2}}\psi dx}=-2{{b}_{c}}\left( \varphi ,{{u}^{2}}\psi  \right)\] Therefore, it yields
${{w}^{\left( s \right)\dagger }}=-{{w}^{\left( s \right)}}$, that is ${{w}^{\left( s \right)}}\in SHer$, then \[\left( \varphi ,{{{\hat{w}}}^{\left( ri \right)}}\psi  \right)=\left( {{{\hat{w}}}^{\left( cl \right)}}\varphi ,\psi  \right)-\left( {{w}^{\left( s \right)}}\varphi ,\psi  \right)=\left( {{{\hat{w}}}^{\left( ri \right)\dagger }}\varphi ,\psi  \right)\]Above all, it yields a conjugate transposes of ${{\hat{w}}^{\left( ri \right)}}$ as follows
 \[{{\hat{w}}^{\left( ri \right)\dagger }}={{\hat{w}}^{\left( cl \right)\dagger }}+{{w}^{\left( s \right)\dagger }}={{\hat{w}}^{\left( cl \right)}}-{{w}^{\left( s \right)}}\ne {{\hat{w}}^{\left( ri \right)}}\]
Thusly,  ${{\hat{w}}^{\left( ri \right)}}$ is a non-Hermitian operator, the proof is finished.
\end{proof}
Note that corollary \ref{c1} tells us that non-Hermitian operator ${{\hat{w}}^{\left( ri \right)}}$ has a complex eigenvalues while Hermitian operator ${{\hat{w}}^{\left( cl \right)}}$ gets the real eigenvalues. Meanwhile, it also implies that the eigenvalues of the non-Hermitian Hamiltonian operator \eqref{a2} are complex form. The non-Hermitian operator \[{{\hat{w}}^{\left( ri \right)}}={{\hat{w}}^{\left( cl \right)}}+{{w}^{\left( s \right)}}\in NHer,~{{\hat{w}}^{\left( cl \right)}}\in Her,~{{w}^{\left( s \right)}}\in SHer\]
and its Hermitian form ${{\hat{w}}^{\left( ri \right)\dagger }}={{\hat{w}}^{\left( cl \right)}}-{{w}^{\left( s \right)}}$, then we obtain
\begin{align}
  & {{\hat{w}}^{\left( ri \right)}}-{{\hat{w}}^{\left( ri \right)\dagger }}=2{{w}^{\left( s \right)}}\in SHer \notag\\
 & {{\hat{w}}^{\left( ri \right)}}+{{\hat{w}}^{\left( ri \right)\dagger }}=2{{{\hat{w}}}^{\left( cl \right)}}\in Her \notag
\end{align}
Conversely, we multiply ${{\hat{w}}^{\left( ri \right)}}$ by $\sqrt{-1}$, then it yields
\[\sqrt{-1}{{\hat{w}}^{\left( ri \right)}}=\sqrt{-1}{{w}^{\left( s \right)}}+\sqrt{-1}{{\hat{w}}^{\left( cl \right)}}\in NHer,\]where $~\sqrt{-1}{{\hat{w}}^{\left( cl \right)}}\in SHer,~\sqrt{-1}{{w}^{\left( s \right)}}\in Her$. As we can see, the non-Hermitian operator ${{\hat{w}}^{\left( ri \right)}}$ has a complex eigenvalues, it means that the imaginary part of its eigenvalues are nonzeros.

\subsection{Proof of the theorem \ref{t2} and theorem \ref{t3}}
\par{\bf Proof of the theorem \ref{t2}}:\par
\begin{proof}
  For the G-dynamics \eqref{a1},
${{\hat{w}}^{\left( cl \right)}}$ is a Hermitian operator making all eigenvalues ${{w}^{\left( q \right)}}$ real stated by the theorem \ref{t1}, for the third statement, it can be proven by formula \[{{\left( \sqrt{-1}{{{\hat{w}}}^{\left( cl \right)}} \right)}^{\dagger }}=-\sqrt{-1}{{\hat{w}}^{\left( cl \right)\dagger }}=-\sqrt{-1}{{\hat{w}}^{\left( cl \right)}}\]
Hence, $\sqrt{-1}{{\hat{w}}^{\left( cl \right)}}$ is skew Hermitian or anti-Hermitian if and only if ${{\hat{w}}^{\left( cl \right)}}$ is a Hermitian operator,  vice versa for the fourth statement.
\end{proof}

\par
{\bf Proof of the theorem \ref{t3}}:\par
\begin{proof}
  Based on the Hermitian operator \eqref{a3}, we can verify
\begin{align}
 \left( \varphi ,\hat{Q}\psi  \right) & =\int{{{\varphi }^{*}}\hat{Q}\psi dx}=\int{{{\varphi }^{*}}\left( 2u{{\psi }_{x}}+\psi {{u}_{x}} \right)dx}=2\int{{{\varphi }^{*}}u{{\psi }_{x}}dx}+\int{{{\varphi }^{*}}{{u}_{x}}\psi dx} \notag\\
 & =2\int{{{\varphi }^{*}}ud\psi }+\int{{{\varphi }^{*}}{{u}_{x}}\psi dx} \notag\\
 & =2\left( u{{\varphi }^{*}}\psi -\int{\psi d\left( {{\varphi }^{*}}u \right)} \right)+\int{{{\varphi }^{*}}{{u}_{x}}\psi dx} \notag\\
 & =-2\int{\psi d\left( {{\varphi }^{*}}u \right)}+\int{{{\varphi }^{*}}{{u}_{x}}\psi dx} \notag
\end{align}
where the integration by parts is used, and ${{\varphi }^{*}}\psi \left| _{\infty } \right.=0$, subsequently, it further leads to the result
\begin{align}
\left( \varphi ,\hat{Q}\psi  \right)  & =-2\int{\psi d\left( {{\varphi }^{*}}u \right)}+\int{{{\varphi }^{*}}{{u}_{x}}\psi dx}=-2\int{\psi \left( {{\varphi }^{*}}_{x}u+{{\varphi }^{*}}{{u}_{x}} \right)dx}+\int{{{\varphi }^{*}}{{u}_{x}}\psi dx}  \notag\\
 & =-2\int{\psi u{{\varphi }^{*}}_{x}dx}-2\int{\psi {{u}_{x}}{{\varphi }^{*}}dx}+\int{{{\varphi }^{*}}{{u}_{x}}\psi dx} \notag \\
 & =-2\int{\psi u{{\varphi }^{*}}_{x}dx}-\int{\psi {{u}_{x}}{{\varphi }^{*}}dx} \notag \\
 & =-\int{\psi {{\left( \hat{Q}\varphi  \right)}^{*}}dx}=-\left( \hat{Q}\varphi ,\psi  \right)  \notag
\end{align}
As a result, the derivation gives the outcome $\left( \hat{Q}\varphi ,\psi  \right)=-\left( \varphi ,\hat{Q}\psi  \right)$, therefore,  the curvature operator $\hat{Q}\in SHer$ is a skew-Hermitian operator satisfying $\hat{Q}^{\dag}=-\hat{Q}$,    then the eigenvalues of curvature operator $\hat{Q}$ are all purely imaginary, we complete the proof.

\end{proof}
Note that the curvature operator can be rewritten as $\hat{Q}={{a}_{r}}\sqrt{-1}{{\hat{w}}^{\left( cl \right)}}$ in terms of the G-dynamics ${{\hat{w}}^{\left( cl \right)}}$, where ${{a}_{r}}=2m/\hbar$, obviously, due to the G-dynamics ${{\hat{w}}^{\left( cl \right)}}$ is Hermitian operator based on the theorem \ref{t1} , according to the theorem \ref{t2}, we can immediately get the conclusion that $\hat{Q}$ is a skew-Hermitian operator. More specifically, it can be deduced by
\[{{\hat{Q}}^{\dagger }}={{\left( {{a}_{r}}\sqrt{-1}{{{\hat{w}}}^{\left( cl \right)}} \right)}^{\dagger }}=-{{a}_{r}}\sqrt{-1}{{\hat{w}}^{\left( cl \right)\dagger }}=-{{a}_{r}}\sqrt{-1}{{\hat{w}}^{\left( cl \right)}}=-\hat{Q}\]
As it shows, it fits the feature of the skew-Hermitian operator.
In fact, it turns out that skew-Hermitian operator $\hat{Q}$ has special status embodied the unique structure of the manifold space.

\subsection{Proof of the theorem \ref{t5}}

\par
{\bf Proof of the theorem \ref{t5}}:\par
\begin{proof}
  It can be directly verified
\begin{align}
 \left( \varphi ,d^\text{2}\psi /dx{}^\text{2} \right) &=\int{{{\varphi }^{*}}\left( d^\text{2}/dx{}^\text{2} \right)\psi dx} =\int{{{\varphi }^{*}}\left( d\left( d\psi /dx \right)/dx \right)dx}=\int{{{\varphi }^{*}}d\left( d\psi /dx \right)} \notag\\
 & ={{\varphi }^{*}}d\psi /dx-\int{\left( d\psi /dx \right)d{{\varphi }^{*}}}=-\int{\left( d\psi /dx \right)d{{\varphi }^{*}}} \notag\\
 & =-\int{\left( d\psi /dx \right)}\left( d{{\varphi }^{*}}/dx \right)dx \notag
\end{align}based on the Hermitian operator \eqref{a3} and the expression
\begin{align}
 ~\left( d^\text{2}\varphi /dx^{2},\psi  \right) &=\int{{{\left( d^\text{2}\varphi /dx^\text{2} \right)}^{*}}\psi dx}=\int{\left( d^\text{2}{{\varphi }^{*}}/dx^\text{2} \right)\psi dx}=\int{\psi d\left( d{{\varphi }^{*}}/dx \right)} \notag\\
 & =\psi d{{\varphi }^{*}}/dx-\int{\left( d{{\varphi }^{*}}/dx \right)d\psi }=-\int{\left( d{{\varphi }^{*}}/dx \right)d\psi } \notag\\
 & =-\int{\left( d{{\varphi }^{*}}/dx \right)\left( d\psi /dx \right)dx} \notag
\end{align}where we have used the fact that if $\varphi$ and $\psi$ are square integrable, they must go to zero at $\pm \infty$, for this reason, it has ${{\varphi }^{*}}d\psi /dx= \psi d{{\varphi }^{*}}/dx=0$ given at $\pm \infty$.  Hence, it leads to the equality
$\int{{{\varphi }^{*}}\left( d^{2}/dx^{2} \right)\psi dx}=\int{{{\left( d^{2}\varphi /dx{}^{2} \right)}^{*}}\psi dx}$, that is $\left( d^{2}\varphi /dx^{2},\psi  \right)=\left( \varphi ,d^{2}\psi /dx^{2} \right)$, therefore, $-\frac{{{\hbar }^{2}}}{2m}\frac{{{d}^{2}}}{d{{x}^{2}}}\in Her$,  and as proven in the theorem  \ref{t2}, $-\sqrt{-1}\hbar {{\hat{w}}^{\left( cl \right)}}\in SHer$, thusly, the one-dimensional motor operator \eqref{a7} ${{\hat{E}}^{\left( w \right)}}\in NHer$ holds.
\end{proof}
Note that the conjugate transposes of the one-dimensional motor operator \eqref{a7} is $${{\hat{E}}^{\left( w \right)\dagger }}=-\frac{{{\hbar }^{2}}}{2m}{{d}^{\text{2}}}/d{{x}^{\text{2}}}+\sqrt{-1}\hbar {{\hat{w}}^{\left( cl \right)}}\in NHer$$ it can be verified as follows
\begin{align}
\left( \varphi ,{{\hat{E}}^{\left( w \right)}}\psi  \right)  &=\int{{{\varphi }^{*}}{{\hat{E}}^{\left( w \right)}}\psi dx}=\int{{{\varphi }^{*}}\left( -\frac{{{\hbar }^{2}}}{2m}{{d}^{\text{2}}}/d{{x}^{\text{2}}}-\sqrt{-1}\hbar {{{\hat{w}}}^{\left( cl \right)}} \right)\psi dx} \notag\\
 & =-\frac{{{\hbar }^{2}}}{2m}\int{{{\varphi }^{*}}{{d}^{\text{2}}}\psi /d{{x}^{\text{2}}}dx}-\sqrt{-1}\hbar \int{{{\varphi }^{*}}{{{\hat{w}}}^{\left( cl \right)}}\psi dx} \notag\\
 & =-\frac{{{\hbar }^{2}}}{2m}\left( {{d}^{\text{2}}}\varphi /d{{x}^{\text{2}}},\psi  \right)+\left( \sqrt{-1}\hbar {{{\hat{w}}}^{\left( cl \right)}}\varphi ,\psi  \right) \notag\\
 & =\left( \left( -\frac{{{\hbar }^{2}}}{2m}{{d}^{\text{2}}}/d{{x}^{\text{2}}}+\sqrt{-1}\hbar {{{\hat{w}}}^{\left( cl \right)}} \right)\varphi ,\psi  \right) \notag\\
 & =\left( {{\hat{E}}^{\left( w \right)\dagger }}\varphi ,\psi  \right) \notag
\end{align}
 then it leads to the outcome of skew-Hermitian operator by a difference operation \[{{\hat{E}}^{\left( w \right)}}-{{\hat{E}}^{\left( w \right)\dagger }}=-2\sqrt{-1}\hbar {{\hat{w}}^{\left( cl \right)}}=\Delta {{\hat{H}}^{\left( ri \right)}}=-2{{\hat{H}}^{\left( \operatorname{clm} \right)}}\in SHer\]

\section{Application of Berry-Keating Hamiltonian operator}
In this section we exemplify the above general formulation of Hamiltonian for Berry-Keating Hamiltonian operator of the form $\hat{H}^{\left( \text{bk}\right)}=-\sqrt{-1}\hbar \left( x\frac{d}{dx}+1/2 \right)$ corresponding to a trial of proving the Hilbert-P\'{o}lya conjecture in one dimension.
Let the Berry-Keating Hamiltonian operator [B-K] $\hat{H}^{\left( \text{bk}\right)}$ be specifically given by
$$\hat{H}^{\left( \text{bk}\right)}=\left( x{\hat{p}}^{\left( cl \right)}+{\hat{p}}^{\left( cl \right)}x \right)/2=-\sqrt{-1}\hbar \left( x\frac{d}{dx}+1/2 \right)\in Her$$ as a Hermitean operator that is constructed for the Hilbert-P\'{o}lya conjecture in the very beginning. As we noticed, the Berry-Keating Hamiltonian operator $\hat{H}^{\left( \text{bk}\right)}=-\sqrt{-1}\hbar \left( x\frac{d}{dx}+1/2 \right)$ is formally similar to the one-dimensional G-dynamics \eqref{a4}.
As an application of the G-dynamics, we plug Berry-Keating's Hamiltonian operator ${{\hat{H}}^{\left( \text{bk}\right)}}$ into the formula  $\hat{w}=-\sqrt{-1}{{\left[ s,\hat{H} \right]}_{QPB}}/\hbar$ of the G-dynamics, and of course, Berry-Keating's Hamiltonian should be completely rewritten as ${{\hat{H}}^{\left( \text{bk}\right)}}=-\sqrt{-1}\hbar {{w}_{0}}\left( x\frac{d}{dx}+1/2 \right)$, where ${{w}_{0}}$ is a constant frequency.  Thusly, then G-dynamics in terms of the Berry-Keating's Hamiltonian operator ${{\hat{H}}^{\left( \text{bk}\right)}}$ in the eigenvalue equation with a function $\psi$ writes
$${{\hat{w}}^{\left( \text{bk}\right)}}\psi=\frac{1}{\sqrt{-1}\hbar }{{\left[ s,{{\hat{H}}^{\left( \text{bk}\right)}}\right]}_{QPB}}\psi$$ by a direct computation, it yields ${{\hat{w}}^{\left( \text{bk}\right)}}\psi={{w}_{0}}xu\psi$.
Hence, it gives the eigenvalue ${{w}_{0}}xu$.
As [I] mentioned, there exists a transformation $u\to x$ such that
$$\hat{Q}=2u\frac{d}{dx}+u_{x}\to \hat{\theta }/2=x\frac{d}{dx}+1/2$$ Actually, the curvature operator can be rewritten $\hat{Q}/2=u\frac{d}{dx}+\frac{1}{2}u_{x}$ that is similar to $\hat{\theta }/2=x\frac{d}{dx}+1/2$ in form, as written together
\[\left\{ \begin{matrix}
   \hat{Q}/2=u\frac{d}{dx}+\frac{1}{2}u_{x}  \\
   \hat{\theta }/2=x\frac{d}{dx}+1/2  \\
\end{matrix} \right.\]
Clearly, we can see how the hidden connection between the Berry-Keating's Hamiltonian operator $${{\hat{H}}^{\left( \text{bk}\right)}}=-\sqrt{-1}\hbar \hat{\theta }/2\in Her$$ and curvature operator $\hat{Q}\in SHer$, we can easily observe the similarity and possible connections, namely, the factor $\hat{\theta }/2=x\frac{d}{dx}+1/2$ appears at here that plays a role to the Berry-Keating's Hamiltonian operator
$${{\hat{w}}^{\left( \text{bk}\right)}}\psi=-{{\left[ s,\hat{\theta }\right]}_{QPB}}\psi/2$$ Notice that ${{\hat{H}}^{\left( \text{bk}\right)}}\in Her$ implies that $\hat{\theta }/2=x\frac{d}{dx}+1/2\in SHer$, for this point, we give a brief proof of the skew-Hermitian operator $\hat{\theta }/2$ as follow,
\begin{align}
 \left( \varphi ,\hat{\theta }\psi /2 \right) & =\int{{{\varphi }^{*}}\left( xd/dx+1/2 \right)\psi dx}=\int{{{\varphi }^{*}}xd\psi }+\frac{1}{2}\int{{{\varphi }^{*}}\psi dx} \notag\\
 & =x{{\varphi }^{*}}\psi-\int{\psi d\left( {{\varphi }^{*}}x \right)}+\frac{1}{2}\int{{{\varphi }^{*}}\psi dx} \notag\\
 & =-\int{\psi {{\varphi }^{*}}dx-}\int{\psi x{{\varphi }_{x}}^{*}dx}+\frac{1}{2}\int{{{\varphi }^{*}}\psi dx}\notag\\
 &=- \int{\psi \left( x{{\varphi }_{x}}^{*}+{{\varphi }^{*}}/2 \right)dx}=- \int{\psi {{\left( \hat{\theta }\varphi /2 \right)}^{*}}dx}\notag\\
 &=-\left( \hat{\theta }\varphi /2,\psi  \right) \notag
\end{align}where  ${{\varphi }^{*}}\psi \left| _{\infty } \right.=0$ has been used.
Hence, we get $\hat{\theta }/2\in SHer$.
For such similar characteristic, it naturally reminds us of the Hilbert-P\'{o}lya conjecture that a Hermitian operator was hidden behind the zeta function non-trivial zeros: their imaginary parts corresponding to the Hermitian operator eigenvalues. Therefore, it implies that the features of the G-dynamics ${{\hat{w}}^{\left( cl \right)}}$ gives us a direction toward the conjecture. This raises a compelling question: is the G-dynamics the Riemann dynamics?  The Hilbert-P\'{o}lya conjecture states the spectral interpretation of the complex zeros of the Riemann zeta function as eigenvalues of a self-adjoint linear operator in some Hilbert space.

In 2017, a paper [B-D] is written by Carl M. Bender, Dorje C. Brody, and Markus P. M\"{u}ller, which builds on Berry's approach to the Hilbert-P\'{o}lya conjecture. ${{\hat{H}}^{\left( \text{bk}\right)}}$ is generalized to a form $\hat{H}^{\left( \text{gbk}\right)}
  =2\hat{\Delta }^{-1}{{\hat{H}}^{\left(\text{bk} \right)}} \hat{\Delta }$ based on the Berry-Keating's Hamiltonian $ \hat{H}^{\left( \text{bk}\right)}$, which they claim satisfies a certain modified versions of the conditions of the Hilbert-P\'{o}lya conjecture,
where $\hat{\Delta }=\mathbb{I}-{{e}^{-\sqrt{-1}{\hat{p}}^{\left( cl \right)}}}$,
it also can be rewritten as
$$\hat{H}^{\left( \text{gbk}\right)}=-\sqrt{-1}\hbar \hat{\Delta }^{-1}{\hat{\theta }}\hat{\Delta }$$
Similarly, taking $\hat{H}^{\left( \text{gbk}\right)}$ into the formula  $\hat{w}=-\sqrt{-1}{{\left[ s,\hat{H} \right]}_{QPB}}/\hbar$ account, then it accordingly obtains  ${{\hat{w}}^{\left( \text{gbk}\right)}}\psi=\frac{1}{\sqrt{-1}\hbar }{{\left[ s,\hat{H}^{\left( \text{gbk}\right)}\right]}_{QPB}}\psi$,  more precisely, it yields
${{\hat{w}}^{\left( \text{gbk}\right)}}\psi =-{{\left[ s,\hat{\Delta }^{-1}{\hat{\theta }}\hat{\Delta } \right]}_{QPB}}\psi={{w}^{\left( \text{gbk}\right)}}\psi$,
where ${{w}^{\left( \text{gbk}\right)}}$ is the eigenvalue of the ${{\hat{w}}^{\left( \text{gbk}\right)}}$ with terms to $\psi$. Here we omit the accurate calculation of the form of the eigenvalue ${{w}^{\left( \text{gbk}\right)}}$.

\section{Conclusions}
The quantum physical observables in the experiment are Hermitian operators, therefore, to understand the characteristics of quantum physical observables, it is particularly important to study the properties of Hermitian operators. All eigenvalues of a self-adjoint operator are real.
With the support of the work of the paper [I], by using the definition of the
Hermitian operator or the self-adjoint operator, we are strictly obedient to its definition and then we have proven that one-dimensional G-dynamics  ${{\hat{w}}^{\left( cl \right)}}$ is a Hermitian operator, this work is very vital to help us ensure its eigenvalues ${{w}^{\left( q \right)}}$ are real only, it reveals that the one-dimensional G-dynamics  ${{\hat{w}}^{\left( cl \right)}}\in Her$ is an physical observable which can be measured in an experiment, it provides the real data for us as proved by experiment, as a result of its Hermiticity, just like other Hermitian operator, it's an objective existence of new dynamic operator, by giving such proof, it strengthens the foundations of the entire  quantum covariant Poisson bracket system. Meanwhile, based on the Hermiticity of one-dimensional G-dynamics  ${{\hat{w}}^{\left( cl \right)}}\in Her$, we also prove that the curvature operator is a skew-Hermitian operator, with the purely imaginary eigenvalues.  At the same time, inspired by the obvious factor $\hat{\theta }/2=x\frac{d}{dx}+1/2\in SHer$, etc, we consider the formula of the G-dynamics to evaluate physical models of the Berry-Keating's Hamiltonian operator and its extensive version as the applications of the G-dynamics. The Hermiticity of one-dimensional G-dynamics ${{\hat{w}}^{\left( cl \right)}}$ allows us to search deeply about the quantum phenomena. As for non-Hermitian Hamiltonians, \[\underbrace{{{{\hat{H}}}^{\left( ri \right)}}}_{\text{non-Hermitian Hamiltonian}}=\underbrace{{{{\hat{H}}}^{\left( cl \right)}}}_{\text{ Hermitian Hamiltonian}}\underbrace{-\underbrace{\frac{{{E}^{\left( s \right)}}}{2}}_{\text{ line curvature}~u}-\underbrace{\sqrt{-1}\hbar {{{\hat{w}}}^{\left( cl \right)}}}_{\text{ G-dynamics }{{{\hat{w}}}^{\left( cl \right)}}}}_{\text{non-Hermitian}}\] which occur naturally as effective interactions, they become more complex and subtler than for Hermitian Hamiltonians.
It turns out that describing natural processes by means of non-Hermitian Hamiltonians is really needed and more complete. As the quantum requirements become more complete, so too does the quantum model.

\section*{Appendix}
In order to help readers navigate this paper and for convenient looking up,   we collect all quantum operators including Hermitian operators (Her), non-Hermitian operators (NHer), and skew-Hermitian operators (SHer) here that appear in this paper, we list them as follows:
 \begin{enumerate}
   \item ${{{\hat{H}}}^{\left( cl \right)}}=-\frac{{{\hbar }^{2}}}{2m}{d}^{2}/d{{x}^{2}}+V(x)\in Her,~~{{\hat{H}}^{\left( g \right)}} ={{{\hat{H}}}^{\left( cl \right)}}-{{E}^{\left( s \right)}}/2\in Her$
   \item ${{\hat{p}}^{\left( cl \right)}}=2mb_{c}d/dx=m{{\hat{v}}^{\left( cl \right)}}\in Her,~~x\in Her$
    \item    ${{\hat{T}}^{\left( ri \right)}}=-\frac{{{\hbar }^{2}}}{2m}\left( {{d}^{2}}/{dx}^{2}+{{u}^{2}} \right)-\sqrt{-1}\hbar {{\hat{w}}^{\left( cl \right)}} =-\sqrt{-1}\hbar{{b}_{c}}{{\hat{Q}}^{\left( c \right)}}\in NHer$
               \item    ${{\hat{Q}}^{\left( c \right)}}={{d}^{2}}/d{{x}^{2}}+{{u}^{2}}+\hat{Q}\in NHer$
   \item ${{{\hat{w}}}^{\left( cl \right)}}=-\sqrt{-1}{{\left[ s,{{\hat{H}}^{\left( cl\right)}}\right]}_{QPB}}/\hbar={{b}_{c}}\left( 2u d/dx+u_{x}\right)={{b}_{c}}\hat{Q}\in Her$
    \item ${{{\hat{w}}}^{\left( ri \right)}}={{\hat{w}}^{\left( cl \right)}}+{{w}^{\left( s \right)}}={{b}_{c}}\left( 2ud/dx+{{u}_{x}}+2{{u}^{2}} \right)\in NHer$
      \item   ${{w}^{\left( s \right)}}=2{{b}_{c}}{{u}^{2}}=-\sqrt{-1}\hbar {{u}^{2}}/m\in SHer$
   \item ${{{\hat{H}}}^{\left( ri \right)}}={{{\hat{H}}}^{\left( cl \right)}}-{{E}^{\left( s \right)}}/2-\sqrt{-1}\hbar {{\hat{w}}^{\left( cl \right)}}={{\hat{H}}^{\left( g \right)}}-{{\hat{H}}^{\left( \operatorname{clm} \right)}}\in NHer$
   \item ${{\hat{H}}^{\left( \operatorname{clm} \right)}}=\sqrt{-1}\hbar {{{\hat{w}}}^{\left( cl \right)}}=\frac{{{\hbar }^{2}}}{2m}\hat{Q}\in SHer$
       \item  $ {{{\hat{H}}}^{\left( \operatorname{rim} \right)}}=\sqrt{-1}\hbar {{{\hat{w}}}^{\left( ri \right)}}={{{\hat{H}}}^{\left( \operatorname{clm} \right)}}+{{E}^{\left( s \right)}}=\frac{{{\hbar }^{2}}}{m}{{u^{2}}}+\sqrt{-1}\hbar {{{\hat{w}}}^{\left( cl \right)}}\in NHer$
        \item    ${{E}^{\left( s \right)}}=\sqrt{-1}\hbar{{w}^{\left( s \right)}}=\frac{{{\hbar }^{2}}}{m}{{u^{2}}}\in Her$
   \item $\hat{Q}={{\hat{w}}^{\left( cl \right)}}/{{b}_{c}}=2u\frac{d}{dx}+u_{x}\in SHer $,~ $\hat{Q}/2=u\frac{d}{dx}+\frac{1}{2}u_{x}\in SHer $
   \item ${{{\hat{E}}}^{\left( w \right)}}=-\frac{{{\hbar }^{2}}}{2m}{d}^{2}/d{{x}^{2}}-\sqrt{-1}\hbar {{\hat{w}}^{\left( cl \right)}}=-\frac{{{\hbar }^{2}}}{2m}\left( {{d}^{2}}/d{{x}^{2}}+\hat{Q} \right)\in NHer$
     \item   $\hat{H}^{\left( \text{bk}\right)}=-\sqrt{-1}\hbar \left( x\frac{d}{dx}+1/2 \right)=-\sqrt{-1}\hbar \hat{\theta }/2\in Her$
        \item  $\hat{H}^{\left( \text{gbk}\right)}=-\sqrt{-1}\hbar \hat{\Delta }^{-1}{\hat{\theta }}\hat{\Delta }\in NHer$
             \item   $\hat{\theta }/2=x\frac{d}{dx}+1/2\in SHer$
   \item  ${{\hat{f}}^{\left( g \right)}}={{{\hat{H}}}^{\left( g \right)}}/\hbar ={{{\hat{H}}}^{\left( cl \right)}}/\hbar -\sqrt{-1}{{w}^{\left( s \right)}}/2\in Her$
\item  ${{{\hat{H}}}^{\left( \operatorname{clm}  \right)}}/\hbar =\sqrt{-1}{{{\hat{w}}}^{\left( cl \right)}}\in SHer$
  \item ${{\hat{f}}^{\left( ri \right)}}={{{\hat{H}}}^{\left( ri \right)}}/\hbar   ={{{\hat{H}}}^{\left( cl \right)}}/\hbar -\sqrt{-1}{{w}^{\left( s \right)}}/2-\sqrt{-1}{{{\hat{w}}}^{\left( cl \right)}}={{\hat{f}}^{\left( g \right)}}-\sqrt{-1}{{\hat{w}}^{\left( cl \right)}}\in NHer$
    \item  ${{\hat{f}}^{\left( w \right)}}={{\hat{E}}^{\left( w \right)}}/\hbar =-\frac{\hbar }{2m}{{d}^{\text{2}}}/d{{x}^{\text{2}}}-\sqrt{-1}{{\hat{w}}^{\left( cl \right)}}\in NHer$
 \end{enumerate}
Meanwhile, gathering all one-dimensional wave equations including the Schr\"{o}dinger equation are listed as follows for comparing the difference and connections.
\begin{description}
  \item[i] $\sqrt{-1}\hbar {{\partial }_{t}}\psi ={{\hat{H}}^{\left( cl \right)}}\psi =-\frac{{{\hbar }^{2}}}{2m}{{\psi }_{xx}}+\psi V\left( x \right)$
  \item[ii] $\sqrt{-1}\hbar {{\hat{w}}^{\left( cl \right)}}\psi ={{\hat{H}}^{\left( \operatorname{clm} \right)}}\psi =\frac{{{\hbar }^{2}}}{2m}\left( 2u{{\psi }_{x}}+\psi {{u}_{x}} \right)$
  \item[iii] $\sqrt{-1}\hbar {{\hat{w}}^{\left( ri \right)}}\psi ={{\hat{H}}^{\left( \operatorname{rim}  \right)}}\psi =\frac{{{\hbar }^{2}}}{2m}\left( 2u{{\psi }_{x}}+\psi {{u}_{x}}+2\psi {{u}^{2}} \right)={{\hat{H}}^{\left( \operatorname{clm} \right)}}\psi+{{E}^{\left( s \right)}}\psi$
  \item[iv] $\sqrt{-1}\hbar \left( {{\partial }_{t}}-{{{\hat{w}}}^{\left( cl \right)}} \right)\psi =-\frac{{{\hbar }^{2}}}{2m}\left( {{\psi }_{xx}}+2u{{\psi }_{x}}+\psi {{u}_{x}} \right)+\psi V\left( x \right)$
  \item[v] $\sqrt{-1}\hbar \left( {{\partial }_{t}}-{{{\hat{w}}}^{\left( ri \right)}} \right)\psi =-\frac{{{\hbar }^{2}}}{2m}\left( {{\psi }_{xx}}+2u{{\psi }_{x}}+\psi {{u}_{x}}+2\psi {{u}^{2}} \right)+\psi V\left( x \right)$
  \item[vi] $\sqrt{-1}\hbar \left( {{\partial }_{t}}+{{{\hat{w}}}^{\left( cl \right)}} \right)\psi =-\frac{{{\hbar }^{2}}}{2m}\left( {{\psi }_{xx}}-2u{{\psi }_{x}}-\psi {{u}_{x}} \right)+\psi V\left( x \right)$
   \item[vii] $\sqrt{-1}\hbar \left( {{\partial }_{t}}+{{{\hat{w}}}^{\left( ri \right)}} \right)\psi =-\frac{{{\hbar }^{2}}}{2m}\left( {{\psi }_{xx}}-2u{{\psi }_{x}}-\psi {{u}_{x}}-2\psi {{u}^{2}} \right)+\psi V\left( x \right)$
\end{description}
Actually, by giving the rigorous Hermitian proof of G-dynamics, we can easily judge how the form of the eigenvalues given by the corresponding eigenvalues equations respectively.  Note that {\bf{i, ii, iii}} can be regarded as independent wave equations while {\bf{iv,v,vi,vii}} are combinational wave equations that can be seen as some kinds of extensions of the Schr\"{o}dinger equation {\bf{i}}.

\section*{References}
\ \ \
\par [B-K]
M V. Berry , J P. Keating. The Riemann zeros and eigenvalue asymptotics [J]. SIAM
Review, 1999, 41(2): 236-266.
\par [G-W] G. Wang. Generalized geometric commutator theory and quantum geometric bracket and its uses. arXiv:2001.08566
\par [B-D]
Carl M. Bender, Dorje C. Brody, Markus P. M\"{u}ller. Hamiltonian for the Zeros of the Riemann Zeta Function. Physical Review Letters, 2017, 118 (13): 130-201.
arXiv:1608.03679
\par [A-G] G. Arfken. Self-Adjoint Differential Equations. in Mathematical Methods for Physicists, 3rd ed. Orlando, FL: Academic Press, 1985, 497-509.
\par [B-L]
L. E. Ballentine. The Statistical Interpretation of Quantum Mechanics, Reviews of Modern Physics, 1970, 42 (4): 358-381
\par [N-G]
N. I. Akhiezer; I. M. Glazman. Theory of Linear Operators in Hilbert Space, Two volumes, Pitman, 1981.
\par [V-M]
V. Moretti. Spectral Theory and Quantum Mechanics:  Mathematical Foundations of Quantum Theories, Symmetries and Introduction to the Algebraic Formulation, Springer-Verlag, 2018.
\par
 [I] J. Whongius. On one-dimensional G-dynamics and non-Hermitian Hamiltonian operators.  arXiv:2108.01947.
\par [E-P]
Eduard. Prugove$\check{c}$ki. Quantum Mechanics in Hilbert Space (2nd ed). Academic Press, 1981.
\par [C-B]
Carl M. Bender.  Making Sense of Non-Hermitian Hamiltonians. Reports on Progress in Physics. 2007, 70 (6): 947-1018. arXiv:hep-th/0703096
\par [A-M]
Ali. Mostafazadeh. Pseudo-Hermiticity versus symmetry: The necessary condition for the reality of the spectrum of a non-Hermitian Hamiltonian. Journal of Mathematical Physics. 2002, 43 (1): 205-214. arXiv:math-ph/0107001

\end{document}